\documentclass[a4paper,fleqn,usenatbib]{mnras}

\usepackage{newtxtext,newtxmath}
\usepackage[T1]{fontenc}
\usepackage{ae,aecompl}
\usepackage{color}

\usepackage{graphicx}
\usepackage{multirow}
\usepackage{xspace}
\usepackage{scalerel}
\usepackage[dvipsnames,usenames,table]{xcolor}
\usepackage{pdflscape}

\newcommand{\trilegal}{\textsc{trilegal}}
\newcommand{\parsec}{\textsc{parsec}}
\newcommand{\colibri}{\textsc{colibri}}

\newcommand{\Msun}{{\rm M}_{\odot}}
\newcommand{\Lsun}{{\rm L}_{\odot}}
\newcommand{\Rsun}{{\rm R}_{\odot}}

\newcommand{\Teff}{T_{\rm eff}}
\newcommand{\aml}{\alpha_{\rm ML}}
\newcommand{\anu}{\alpha_{\nu}}

\newcommand{\xc}{X_{\rm C}}
\newcommand{\xo}{X_{\rm O}}
\newcommand{\co}{{\rm C}/{\rm O}}

\newcommand{\wjk}{W_{\scaleto{\rm J,K}{4.5pt}}}
\newcommand{\wrp}{W_{\scaleto{\rm B,R}{4.5pt}}}
\newcommand{\ks}{K_{\rm s}}
\newcommand{\grp}{G_{\scaleto{\rm RP}{4.5pt}}}
\newcommand{\gbp}{G_{\scaleto{\rm BP}{4.5pt}}}

\newcommand{\gaia}{\textit{Gaia}}

\newcommand{\Rb}{R_{\rm b}}
\newcommand{\Pb}{P_{\rm b}}
\newcommand{\Rs}{R_{\rm s}}
\newcommand{\Ps}{P_{\rm s}}
\newcommand{\kb}{k_{\rm b}}
\newcommand{\Mb}{M_{\rm b}^k}
\newcommand{\gamone}{\gamma_1^k}
\newcommand{\gamtwo}{\gamma_2^k}

\defcitealias{Trabucchi_etal_2019}{Paper I}

\title[Nonlinear fundamental pulsation of LPVs]
{Modelling Long-Period Variables -- II.\\
Fundamental mode pulsation in the nonlinear regime}

\author[M. Trabucchi et al.]{
Michele Trabucchi,$^{1,2}$\thanks{Contact e-mail: \href{mailto:michele.trabucchi@unipd.it}{michele.trabucchi@unige.ch}}
Peter R. Wood$^{3}$,
Nami Mowlavi$^{1}$,
Giada Pastorelli$^{2,4}$,
\newauthor
Paola Marigo$^{2}$,
L{\'e}o Girardi$^{5}$,
Thomas Lebzelter$^{6}$
\\
$^{1}$Department of Astronomy, University of Geneva, Ch. des Maillettes 51,
1290 Versoix, Switzerland \\
$^{2}$Dipartimento di Fisica e Astronomia, Universit\`a di Padova,
Vicolo dell'Osservatorio 2, I-35122 Padova, Italy \\
$^{3}$Research School of Astronomy and Astrophysics,
Australian National University, Canberra, ACT 2611, Australia \\
$^{4}$STScI, 3700 San Martin Drive, Baltimore, MD 21218, USA \\
$^{5}$Osservatorio Astronomico di Padova - INAF, Vicolo dell'Osservatorio 5, I-35122 Padova, Italy \\
$^{6}$University of Vienna, Department of Astrophysics, Tuerkenschanzstrasse 17, 1180 Vienna, Austria
}

\date{Accepted XXX. Received YYY; in original form ZZZ}

\pubyear{2020}

\begin{document}
\label{firstpage}
\pagerange{\pageref{firstpage}--\pageref{lastpage}}
\maketitle

\begin{abstract}
Long-period variability in luminous red giants has several promising applications, all of which require models able to accurately predict pulsation periods. Linear pulsation models have proven successful in reproducing the observed periods of overtone modes in evolved red giants, but they fail to accurately predict their fundamental mode periods. Here, we use a 1D hydrodynamic code to investigate the long-period variability of M-type asymptotic giant branch stars in the nonlinear regime. We examine the period and stability of low-order radial pulsation modes as a function of mass and radius, and find overtone mode periods in complete agreement with predictions from linear pulsation models. In contrast, nonlinear models predict an earlier onset of dominant fundamental mode pulsation, and shorter periods at large radii. Both features lead to a substantially better agreement with observations, that we verify against OGLE and \gaia\ data for the Magellanic Clouds. We provide simple analytic relations describing the nonlinear fundamental mode period-mass-radius relation. Differences with respect to linear predictions originate from the readjustment of the envelope structure induced by large-amplitude pulsation. We investigate the impact of turbulent viscosity on linear and nonlinear pulsation, and probe possible effects of varying metallicity and carbon abundance.
\end{abstract}




\section{Introduction}
\label{sec:Introduction}

Low- to intermediate-mass stars approach the end of their evolution through the asymptotic giant branch (AGB) phase, during which they develop large-amplitude variability characterized by periodicities of tens to hundreds of days. These objects are known as long-period variables (LPVs), a rather generic term encompassing distinct evolutionary phases with a variety of observed features and physical properties.

Mira variables, the most spectacular among the LPVs, are AGB stars thought to pulsate in the fundamental radial mode, and follow a well known period-luminosity (PL) relation \citep[e.g.][]{Feast_1984,Feast_etal_1989}. Several other PL relations of LPVs have been discovered in the Large Magellanic Cloud (LMC) using data from the MACHO survey \citep{Wood_etal_1999,Wood_2000}, and are interpreted as resulting from pulsation in overtone modes \citep[e.g.][]{Wood_2015,Trabucchi_etal_2017} or ellipsoidal light variations in binary systems \citep{Soszynski_etal_2004_ellipsoidal}. One sequence is formed by periods systematically longer that those of Miras, and cannot be reconciled with the same kind of stellar oscillation invoked to explain other sequences. They are called long secondary periods (LSPs), and their nature is currently unknown \citep[e.g.][and references therein]{Nicholls_etal_2009,Takayama_Ita_2020}.

The PL relations of LPVs have recently been drawing attention due to their potential as distance indicators \citep{Whitelock_2013,Rau_etal_2019,Yuan_etal_2017M33,Yuan_etal_2017LMC,Yuan_etal_2018,Huang_etal_2018,Huang_etal_2020}. Indeed, Mira variables are easy to find and identify: they are intrinsically bright, have large photometric amplitudes, and are common stars, found in rather diverse astrophysical environments. For similar reasons, LPVs are often used to characterize stellar populations over a wide range of ages and metallicities \citep[e.g.][]{Lebzelter_etal_2018}, and have the potential to be used as age indicators \citep[e.g.][]{Feast_Whitelock_2000,Feast_etal_2006,Grady_etal_2019}.

The strong dependence of pulsation properties upon stellar structure makes for an excellent tool to infer otherwise elusive stellar parameters such as radii and current masses. This is especially true in the case of multi-periodicity, a common feature in LPVs, that offers a way to obtain stringent constraints on modelling by simultaneously satisfying several observed variability requirements. In this respect, many LPVs (especially at RGB luminosities) show properties
that are somewhat akin to solar-like oscillating red giants, \citep[e.g.][]{Dziembowski_Soszynski_2010,Mosser_etal_2013,Stello_etal_2014} suggesting the possibility to apply the successful investigation tools of asteroseismology to more evolved stars.

Pulsation is of special interest in the context of AGB stars, as it plays a crucial role in the mass-loss process that leads to their death \citep[see e.g. the review by][]{Hofner_Olofsson_2018}. In brief, large-amplitude pulsation drives energetic shock waves through the atmosphere, generating favourable physical conditions for the condensation of dust grains. Stellar radiation then transfers momentum to dust (and, via dynamical coupling, to the gas), driving intense outflows from the surface. The correlation of pulsation in different modes with mass loss and dust formation has been explored, e.g., by \citet{Lebzelter_etal_2006}, who linked the change of dominant dust species of AGB stars in 47 Tuc with the switch from overtone to fundamental pulsation, and more recently by \citet{McDonald_Trabucchi_2019}, who examined the signature of the transition between distinct mass-loss rate regimes in the PL relations of LPVs in the LMC.

The importance of a good understanding of mass loss on the AGB can hardly be overstated. On one side, it determines AGB lifetimes, a knowledge of which is necessary to accurately quantify the important contribution of these bright stars
to the integrated light of galaxies and their spectral energy distribution \citep[e.g.][]{Maraston_2005}. On the other side, it can be an important contributor in the chemical evolution of galaxies, as it pollutes the interstellar medium with ejecta that have been enriched in nucleosynthesis products by repeated third dredge-up events, and by CNO-cycle burning at the bottom of the convective envelope in the most massive AGB stars \citep{Kobayashi_etal_2011,Karakas_Lattanzio_2014}.

LPVs are thus critical ingredients in several fields of astrophysical research, and efforts are being made to establish solid connections between their pulsation features and their physical and evolutionary properties, in order to fully exploit their potential. This is especially important in view of the large volume of stellar variability data expected from current and upcoming large-scale surveys, such as \gaia\ \citep{GaiaCollaboration_2018} and the Rubin Observatory Legacy Survey of Space and Time \citep[LSST, ][]{Ivezic_LSST_2109}. Some recent results include evidence for the Mira PL relation being independent of metallicity \citep{Goldman_etal_2019}, and the exploitation of the \gaia\ Data Release 2 (DR2) catalog of LPV candidates \citep{Mowlavi_etal_2018} to characterize the mass dependence of PL relations of LPVs in the Magellanic Clouds \citep{Lebzelter_etal_2019}. On the theoretical side, effort has been devoted to the implementation of descriptions of time-dependent convection to model its coupling with pulsation in one-dimensional models \citep[e.g.][and references therein]{Olivier_Wood_2005,Xiong_etal_2018}, while the innovative three-dimensional ``star-in-a-box'' models of AGB stars by \citet{Freytag_etal_2017} naturally develop pulsation with periods compatible with observations.

In the previous work of this series \citep[][hereafter \citetalias{Trabucchi_etal_2019}]{Trabucchi_etal_2019}, we have addressed the theoretical study of LPVs by computing a large grid of linear 1D models of non-adiabatic radial pulsation. This allowed both for the derivation of easy-to-use tools and prescriptions to predict variability features as a function of global stellar parameters, and to combine them with synthetic stellar population simulations to test current pulsation models against the benchmark of resolved stellar populations in the Magellanic Clouds. This approach confirmed the validity of linear predictions for overtone mode pulsation \citep{Trabucchi_etal_2017}, while showing how they systematically overestimate the period of fundamental mode pulsation. In the latter, the amplitude of radial displacement of the stellar layers during a pulsation cycle becomes very large, hence the breakdown of the linear approximation. Miras and related fundamental mode pulsators are possibly the type of LPVs with the most promising applications, but most of the available prescriptions used to describe their variability are based on linear models, intrinsically inadequate for this task \citep{Trabucchi_etal_2020_ViennaVarStar}.

Nonlinear hydrodynamic simulations are naturally more suited to describe large-amplitude pulsation \citep{YaAri_Tuchman_1996,Lebzelter_Wood_2005,Ireland_Scholz_Wood_2008,Kamath_etal_2010}, and have been employed in several works to study LPVs [see e.g. \citet{Olivier_Wood_2005}]. Yet, all such studies are limited to a small number of models, while, to date, there exists no systematic investigation of nonlinear pulsation in luminous red giants as a function of global stellar parameters. As a result, it is still unclear how different are the predictions coming from linear and nonlinear models. In the present work, we address this kind of study with the help of a one-dimensional hydrodynamic code that includes a time-dependent treatment of convection. This gives us the opportunity of investigating in detail how pulsation is affected by the dissipation of kinetic energy due to turbulent viscosity, another open issue.

This paper is structured as follows. In Sect.~\ref{sec:ModelsAndParameters} we introduce the main features of our methodology and the range of stellar parameters covered in this study. In Sect.~\ref{sec:StaticEnvelopesAndLinearStability} we describe the calculation and properties of the set of reference linear pulsation models. The effect of turbulent viscosity on linear pulsation is also discussed. The calculation and processing of nonlinear models are described in Sect.~\ref{sec:NonlinearModels}. Sect.~\ref{sec:Results} is dedicated to the discussion and interpretation of the results, and to the modelling of the period-mass-radius relation of nonlinear fundamental mode pulsation. We compare our models with observations in Sect.~\ref{sec:ComparisonWithObservations}, while Sect.~\ref{sec:Conclusions} is dedicated to conclusions.

\section{Models and parameters}
\label{sec:ModelsAndParameters}

Firstly, we compute a set of static envelope models representative of M-type (O-rich) AGB stars, and examine their linear and nonlinear radial pulsation. The set is constructed by varying the stellar parameters mass $M$ (generally different than the initial mass $M_{\rm i}$) and luminosity $L$ while we keep fixed chemical composition (metallicity $Z$, hydrogen mass fraction $X$, and number ratio $\co$ of carbon-to-oxygen atoms at the surface). The chosen metallicity ($Z=0.006$) is roughly representative of the young and intermediate-age populations in the Small and Large Magellanic Clouds (SMC and LMC), our reference targets for comparison with observations. The mixing length parameter $\aml$ (mixing length in units of pressure scale height) is also varied, so to obtain a few different values of effective temperature $\Teff$ (i.e. of radius $R$) at fixed mass and luminosity. The mass of the core is artificially increased with luminosity through an analytic relation derived from evolutionary models (see Sect.~\ref{ssec:StaticModels}).

The relevant parameters in the models and corresponding values are summarized in Table~\ref{tab:ModelParameters}, and cover a subset of the grid of \citetalias{Trabucchi_etal_2019}. The only exception is that we explore the effect of turbulent viscosity, described by the free parameter $\anu$, which was fixed to $\anu=0$ in our previous work\footnote{
    Test calculations suggest that a nonzero value of $\anu$ favours convergence of the hydrodynamic code, hence the choice $\anu=10^{-4}$. This should be considered effectively equivalent to choosing $\anu=0$ (no turbulent viscosity), there being negligible differences between the respective results.
}. Increasing $\anu$ makes more efficient the dissipation of pulsational kinetic energy by convective turbulence, and is expected to stabilize pulsation (both in the linear and nonlinear regimes) and to reduce the amplitude of pulsation in nonlinear models. The present work represents the first systematic study of these effects.

It is worth clarifying the meaning of amplitude adopted throughout this paper. In each mass zone of nonlinear models, all relevant physical properties display some degree of oscillation, the amplitude of which can be defined as the difference between maximum and minimum values during the pulsation cycle. Here, we will usually consider the peak-to-peak amplitude of radial displacement of the optical surface (where the Rosseland mean optical depth is $\sim2/3$). While this is correlated with the variation of bolometric luminosity, the latter does not necessarily have the same amplitude as that in observed finite pass bands such as VIJHK. These amplitudes are determined by radiative processes in the atmospheric layers that are not dealt with in the present models.

\begin{table}
    \centering
    \caption{
        Parameters varied in the computation of pulsation models, and corresponding values. The exact boundaries of the luminosity range depend on other parameters.
        }\label{tab:ModelParameters}
    \begin{tabular}{c|c}
        Parameter & Values \\
        \hline
        $Z$ & 0.006 \\
        $X$ & 0.7 \\
        $\co$ & 0.55 \\
        $\aml$ & 1.5, 2.0, 2.5 \\
        $\anu$ & $10^{-4}$, 0.05, 0.1, 0.2 \\
        $M/\Msun$ & 0.6, 1.0, 1.6, 2.6, 4.4, 7.0 \\
        $\log(L/\Lsun)$ & [2.5, 5.0], step: 0.01 \\
    \end{tabular}
\end{table}

\section{Linear pulsation models}
\label{sec:StaticEnvelopesAndLinearStability}

We used the codes described in \citet[][and references therein]{Wood_Olivier_2014} to compute static envelope models and to examine their linear pulsation properties. We briefly summarise the procedure (we refer to \citetalias{Trabucchi_etal_2019} for more details) and examine the results, which will be used as reference for the analysis of nonlinear models.

\subsection{Static models}
\label{ssec:StaticModels}

Spherically symmetric static models are obtained by integrating the envelope structure from a layer sufficiently high in the atmosphere down to a rigid\footnote{
    Due to the strong density contrast between stellar core and envelope, the amplitude of pulsation becomes negligible near the bottom of the envelope, i.e. the core and envelope are dynamically decoupled. This justifies ignoring the core region, a common approach in modelling stellar pulsation.
} core of very small radius ($0.15\,\Rsun$) for which mass and output luminosity are provided as boundary conditions. This innermost region, not modelled, would encompass the actual CO core produced during core He-burning stages, as well as the nuclear-burning shells. The whole stellar luminosity is assumed to be generated within the core and to be constant throughout the envelope. To ensure that static models are representative of AGB envelopes, a core mass-luminosity relation (CMLR) based on stellar evolutionary models is assumed, namely the mean value between the two CMLRs presented in \citetalias{Trabucchi_etal_2019} (Eq.~5). Additional boundary conditions are the total mass and the chemical composition, which is assumed to be homogeneous due to efficient convective mixing. Convection is treated by means of the usual mixing length theory, and the mixing length parameter $\aml$ is used as a control parameter to change the effective temperature. Radiative Rosseland mean opacities as a function of density and temperature and fully consistent with the envelope metal mixture [based on a solar-scaled mixture derived from \citet{Caffau_etal_2011}, but with additional changes in CNO abundances as exemplified in Sect.~\ref{ssec:EffectOfChemicalComposition}] are supplied through external tables, including up-to-date molecular contributions \citep{Marigo_Aringer_2009}.

It is worth pointing out that, given the assumptions made, our static models are not totally accurate descriptions of the envelopes of the most massive AGB stars, which are expected to undergo hot bottom burning (HBB). These stars are overluminous with respect to expectations from the CMLR. In particular, for a given mass and luminosity, a smaller core mass should be assumed to describe a HBB star. In \citetalias{Trabucchi_etal_2019}, linear pulsation models were computed with two different CMLRs in order to account, at least partially, for such situations, as well as for the differences associated with the occurrence of thermal pulses. It was found that varying the core mass and radius has little impact on linear pulsation properties. Here, we assume this to be the case for nonlinear models as well. A detailed study of the role of core parameters in nonlinear pulsation, especially in high-mass AGB stars, as well as of the possible impact of HBB, is nonetheless desirable.

\subsection{Stability}
The linear stability analysis involves solving the linearized equations of non-adiabatic, radial oscillations about an equilibrium configuration, which is described by static models. Relevant physical properties are assumed to depend on time in the form $\exp(\omega t)$ with a complex frequency $\omega=\omega_{\rm R}+\mathbf{i}\omega_{\rm I}$, and time-dependent convection is treated as described in \citet{Fox_Wood_1982}. The code searches for solutions (eigenfrequencies and wave functions) by exploring the complex frequency plane starting from a user-defined region. For each envelope model, we compute the five lowest-order solutions, or pulsation modes, described by their period $P_n=2\uppi/\omega_{{\rm I},n}$ and growth rate $GR_n=\exp(2\uppi\omega_{{\rm R},n}/\omega_{{\rm I},n})-1$. The radial order $n=0$ corresponds to the fundamental mode (FM), while $n=1$ is the first overtone mode (1OM), and so on. The growth rate represents the fractional increase in the amplitude of surface radial displacement per pulsation cycle, and is an indication of the degree to which a mode is stable or excited. It is assumed that modes with a positive value of the growth rate are (linearly) excited, while a stable mode has negative growth rate. The mode with the largest growth rate for a given envelope model is identified with the ``dominant'' mode, which is expected to have the strongest signature in the observed light curve of a pulsating star.

\subsection{Model sequences}
To ensure convergence, it is necessary to define a suitable region in the complex frequency plane where the code will begin to search for solutions. This is less of an issue at relatively low luminosity, where pulsation is almost adiabatic ($|\omega_{\rm R}|\simeq0$), but is crucial in bright models that undergo strongly non-adiabatic pulsation. It is thus convenient to organize the computation of envelopes in ``model sequences'', or ``luminosity sequences''. These are one-parameter families of models in which mass and composition are fixed (as well as $\aml$ and $\anu$), while the luminosity is increased by small steps. After the first few steps, eigenvalues are searched for around the complex frequency extrapolated from previous models. Convergence issues might still arise, especially towards the high-luminosity end of model sequences. In such cases a few iterations are attempted by expanding the search region, and if convergence is not reached the model is skipped and computation moves towards the next one along the sequence. This can lead to small gaps in the sequences (cf. Sect.~\ref{ssec:TurbulentViscosityLinearPulsation}), that do not affect the results of our study.

Note that the effective temperature changes along a sequence, and so does the core mass (through the CMLR), while at the same time the envelope mass decreases. As long as the latter is not too small, a luminosity sequence is representative of the Hayashi line for given mass, composition, and model parameters. This facilitates the analysis of results as, to some extent, luminosity sequences mimic the average evolution of a star along the AGB. Nonetheless, luminosity sequences should not be confused with evolutionary tracks, as they lack a description of many crucial processes (nuclear reactions, mass-loss, stellar winds, thermal pulses and dredge-up events). Even though model sequences are constructed by varying luminosity, in the context of pulsation they are more conveniently parametrized by the value of the surface radius. In fact, varying any parameter of Table~\ref{tab:ModelParameters} (except for mass) affects pulsation periods indirectly mainly by changing the radius of the model \citepalias[see the discussion in ][]{Trabucchi_etal_2019}. In other words, the relation between period, mass, and radius is independent of other stellar and model parameters to a high degree of approximation. For this reason, throughout this paper, the evolution of pulsation along model sequences is usually examined as a function of radius.

\begin{figure}
    \includegraphics[width=.99\columnwidth]{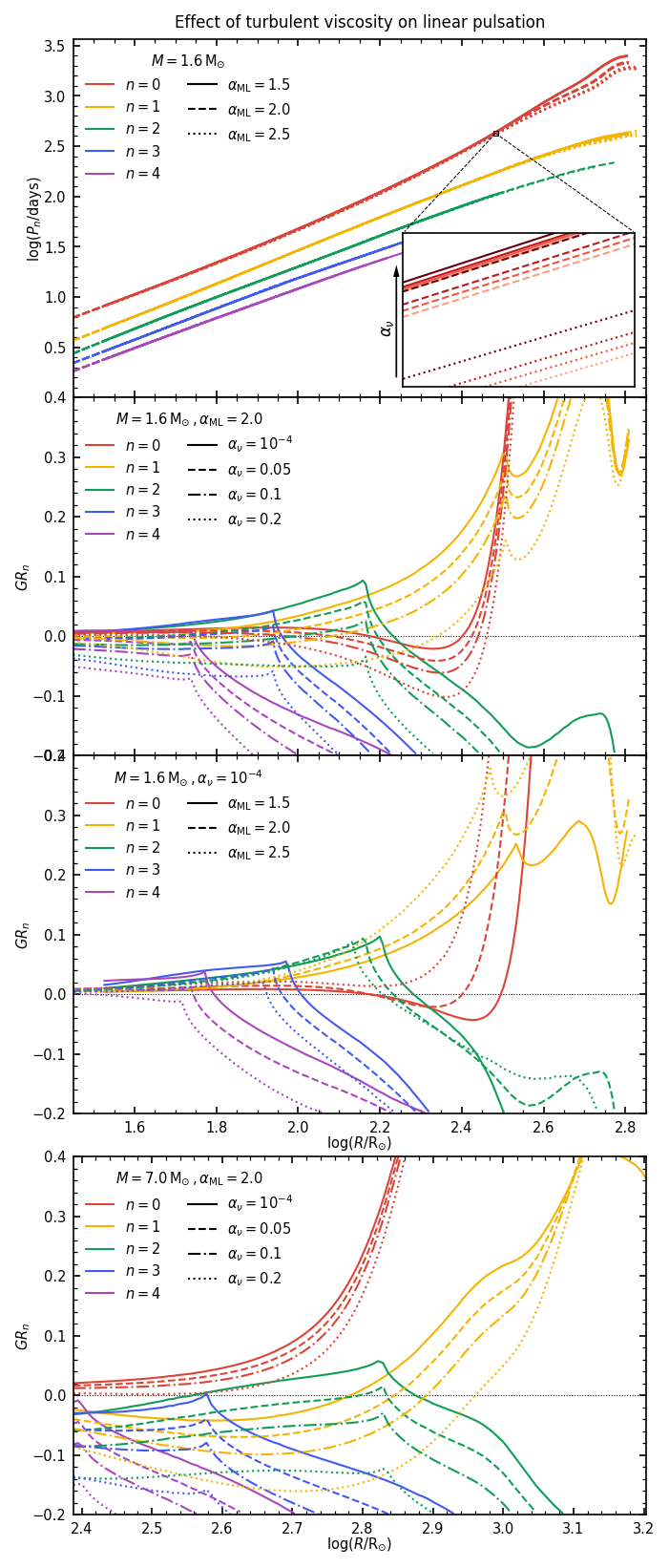}
    \caption{
    Linear periods and growth rates as a function of radius for selected sequences, colour coded according to pulsation modes, showing the effect of varying $\aml$ and $\anu$. Sequences with $M=1.6\,\Msun$ are shown in the top three panels, while the bottom panel shows the case $M=4.4\,\Msun$. In the top panel, the line style indicate the value of $\aml$, while darker tones indicate larger values of $\anu$ (visible only in the enlarged inset panel, showing an enlarged view of the fundamental mode sequence). Other panels show the effect on growth rates of varying either $\aml$ or $\anu$, with distinct line styles corresponding to distinct values as indicated in the legend of each panel. The values of fixed parameters are also indicated in each panel.
    }
    \label{fig:linear_anueffects_example}
\end{figure}

\begin{figure*}
    \includegraphics[width=\textwidth]{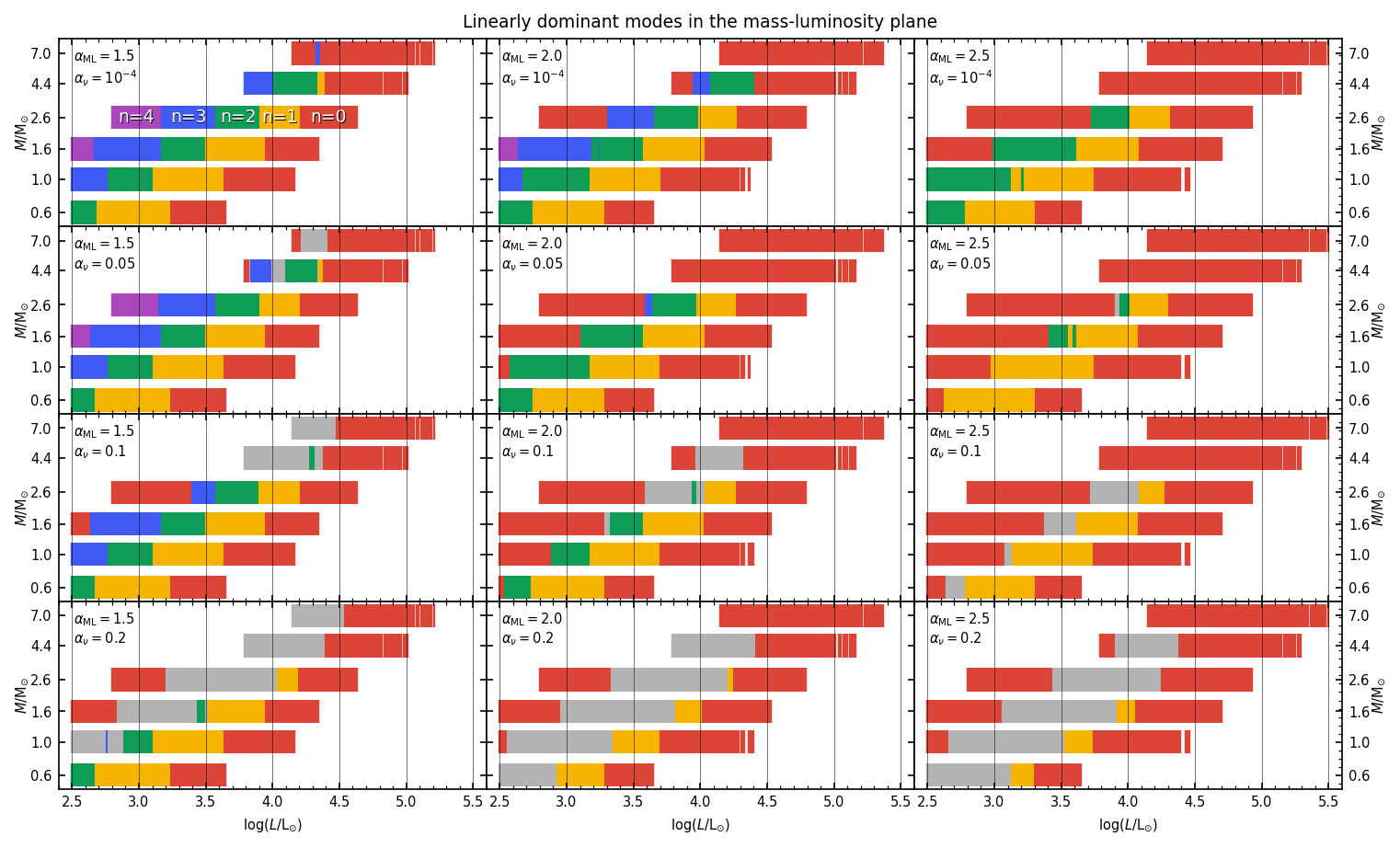}
    \caption{
    Each panel shows the regions where, in the mass-luminosity plane, distinct pulsation modes are dominant according to linear predictions. Red patches indicate regions where the fundamental mode is dominant, and similarly for the overtone modes, using the same colour code as in Fig.~\ref{fig:linear_anueffects_example} (summarized in the top-left panel). Grey patches represent regions where all modes are stable. The mixing length parameter $\aml$ increases for panels from left to right, while the turbulent viscosity parameter $\anu$ increases for panels from top to bottom. Note that the scale is logarithmic along the horizontal axis.
    }
    \label{fig:linearDomPatches_ML}
\end{figure*}

\subsection{Turbulent viscosity and linear pulsation}
\label{ssec:TurbulentViscosityLinearPulsation}

We consider first the effect of turbulent viscosity on linear periods. Since it is found to be similar for all explored values of mass, we show only the case $M=1.6\,\Msun$ (top panel of Fig.~\ref{fig:linear_anueffects_example}). There, the period-radius (PR) relation for different combinations of $\aml$ and $\anu$ is displayed (in this paper, $R$ is the radius where the Rosseland mean optical depth is $\tau_{\rm R}\simeq2/3$). Varying either parameter leads to small changes. Variations of $\aml$ change the radius at which models deviate from their Hayashi line \citepalias[cf. fig.~16 of][]{Trabucchi_etal_2019}, hence the period differences at large radii. In comparison, the effect of changing $\anu$ over the range of likely values is small, and can be appreciated only in the highly enlarged inset panel of Fig.~\ref{fig:linear_anueffects_example}. We conclude that linear periods are almost independent of turbulent viscosity.

Linear stability is more sensitive to the value of $\aml$ and $\anu$. Let us consider first the effect of turbulent viscosity on growth rates for the $1.6\,\Msun$ sequences, as displayed in the second panel from the top of Fig.~\ref{fig:linear_anueffects_example} (having fixed $\aml=2.0$ for clarity). One can recognize the pattern described in \citetalias{Trabucchi_etal_2019}, with overtone growth rates gradually increasing with radius until their period reaches the acoustic cut-off, and they rapidly become stable. The FM growth rate shows only a mild increase at first, and it even decreases for a while, bringing the FM to a temporary stabilization, before growing very rapidly. Qualitatively, this pattern is not altered by varying $\anu$: the growth rates of all modes are reduced, but the effect is stronger for the overtones, especially at the smallest radii. At $\log(R/\Rsun)\lesssim1.9$, increasing $\anu$ from $10^{-4}$ to 0.2 leads to a decrease of about 0.02 in the FM growth rate. To achieve the same change for overtone modes it is sufficient to increase $\anu$ to 0.05. Therefore, increasing turbulent viscosity leads to efficient suppression of overtone modes and favours the dominance of the FM.

Increasing $\aml$ has the effect of decreasing the growth rate at a given $R$ for the 3OM and 4OM, but actually increases the growth rate off the 1OM and FM, while leaving the 2OM with mixed effects (third panel from the top in Fig.~\ref{fig:linear_anueffects_example}). For large enough values of $\aml$, the temporary stabilization of the FM is lifted, so that its growth rate is essentially flat until it reaches the point where it increases very steeply. The effect is more pronounced in the large-mass models, with FM growth rates increasing monotonically with radius, as exemplified in the bottom panel of Fig.~\ref{fig:linear_anueffects_example}. This panel also shows the relatively small effect on the FM growth rate of an increase of $\anu$.

A more general picture is given in Fig.~\ref{fig:linearDomPatches_ML} which shows the regions where distinct modes are dominant in the mass-luminosity plane for different combinations of $\aml$ and $\anu$. Narrow white strips in the FM-dominated areas, at large $M/L$, correspond to the few models skipped due to convergence issues while computing luminosity sequences. The fact that overtone modes are suppressed for large values of $\aml$ and $\anu$ is very evident. Since overtone modes (at least up to the 3OM) are observed at AGB luminosities \citep{Trabucchi_etal_2017,Yu_etal_2020}, this suggests that relatively low values of $\anu$ and $\aml$ are appropriate, at least at low $L/M$. At higher luminosities, no combination of $\aml$ and $\anu$ can be ruled out as in all the explored cases the fundamental mode becomes eventually dominant. It should also be noted that there is no guarantee that these parameters should have constant values, while there is actually some evidence of the contrary. For instance, \citet{Lebzelter_Wood_2016} used an earlier version of the linear pulsation code employed here to model the AGB stars in the LMC cluster NGC 1846, and found that $\aml$ has to increase with luminosity in order to simultaneously reproduce the observed photometry and variability. \citet{Ireland_Scholz_Wood_2011} modelled the variability of a few nearby Miras using values of $\anu$ in the range 0.25-0.32, which, in view of our results, suggests that $\anu$ should also increase with luminosity.

The mild sensitivity of the FM growth rates on turbulent viscosity allow it to become dominant, even though weakly excited, at the low-luminosity end of the model sequences when overtone modes are suppressed. Hence linear models predict that the FM can be dominant at low $L/M$, with small growth rates, and later at high luminosities and with very large growth rates, while overtone modes are dominant between these two regimes. If the increase in $\anu$ is such that the 1OM and 2OM become stable, and if the fundamental mode undergoes temporary stabilization, no pulsation occurs in between.

\section{Nonlinear pulsation models}
\label{sec:NonlinearModels}

\subsection{Computation and general features}
\label{ssec:ComputationAndGeneralFeatures}

To probe the nonlinear pulsation of each envelope model along the luminosity sequences we use the 1D hydrodynamic code described in \citet{Wood_1974}, with updates described in \citet{Keller_Wood_2006}. Briefly, the code solves the equations of nonlinear, radial stellar pulsation incorporating a time-dependent mixing length theory of convection for energy transport and damping of pulsation by turbulent viscosity. Pulsation is treated as an initial value problem, and the equations in difference form are solved at a time $t_n$ given a model at time $t_{n-1}=t_n-\Delta t_n$. A static envelope model is used at $t_0$ as initial condition.

\begin{landscape}
    \begin{figure}
    \centering
        \includegraphics[height=.42\textheight]{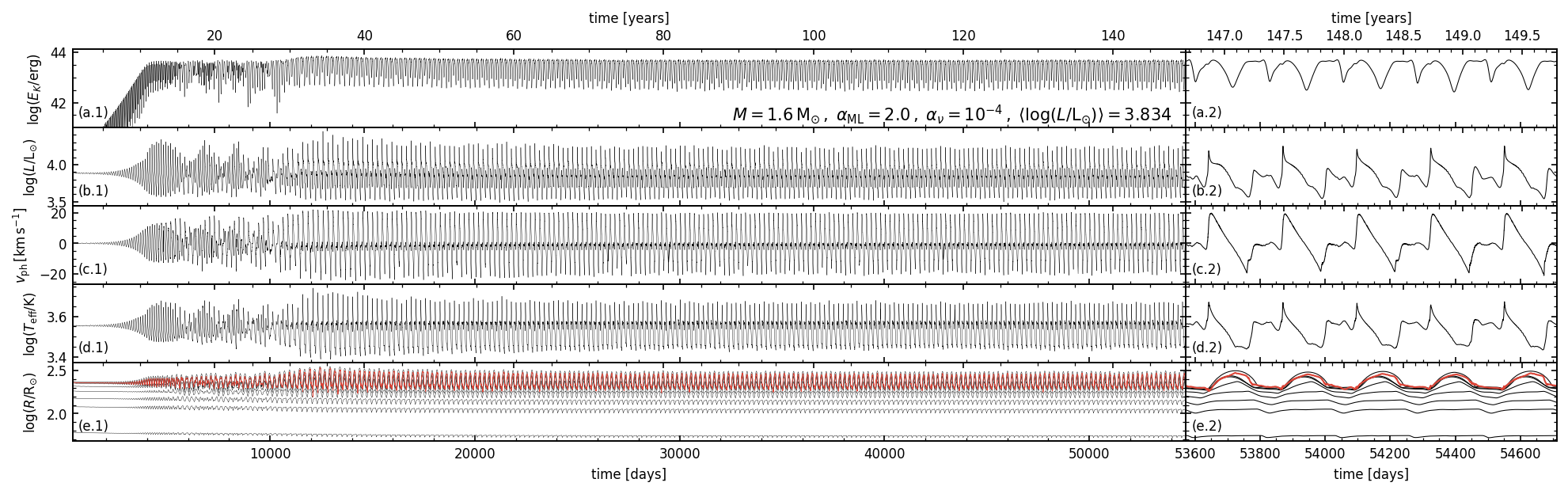}
        \caption{
        Properties of the nonlinear time series at $\log(L/\Lsun)=3.85$ from the sequence with $M=1.6\,\Msun$, $\aml=2.0$, $\anu=10^{-4}$. Panels from top to bottom show the time evolution of kinetic energy, total bolometric luminosity, effective temperature, surface velocity, and radius. In the bottom panel the the radius of several mass zones is shown, as well as the radius of the optical surface (where the Rosseland mean optical depth is $\tau_{\rm R}=2/3$, red line). The shift of the optical surface between different mass zones during a pulsation cycle is evident. Panels on the right column show an enlarged view of the last five pulsation cycles.
        }
        \label{fig:nlSTS_example1}
        \includegraphics[height=.42\textheight]{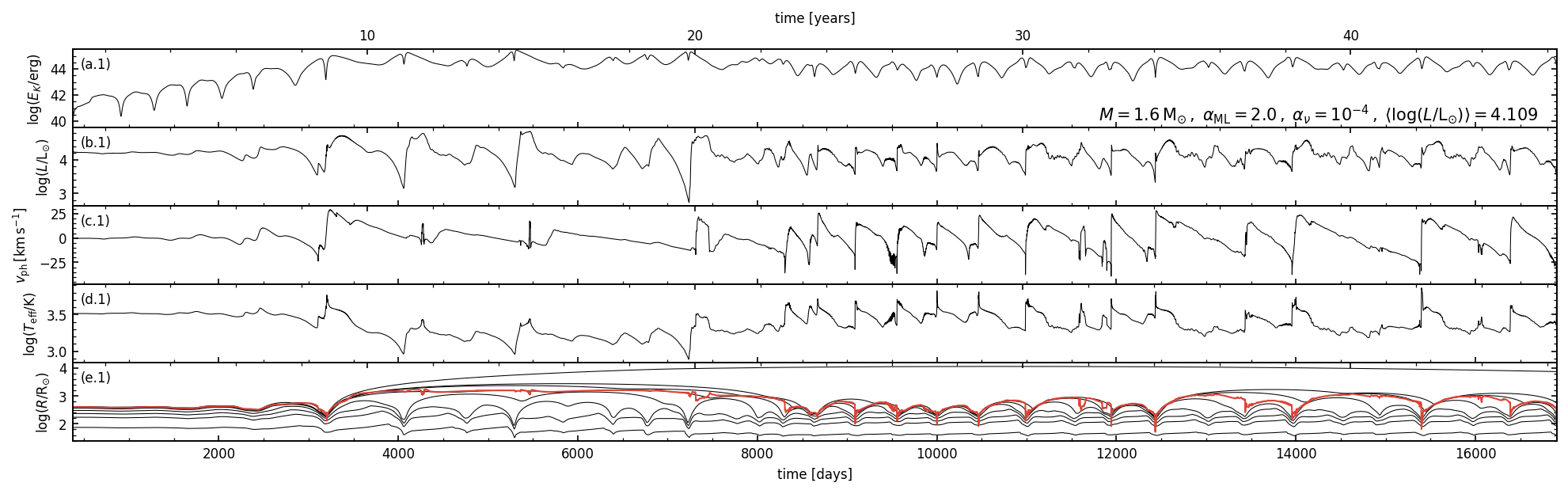}
        \caption{
        Same as Fig.~\ref{fig:nlSTS_example1}, but for $\log(L/\Lsun)=4.18$, representing an example of ``flagged'' time series, whose estimated dominant period might be unreliable.
        }
        \label{fig:nlSTS_example2}
    \end{figure}
\end{landscape}

At low $L/M$, models tend to be stable. As $L/M$ increases, the models become unstable to pulsation: nonlinear pulsation grows from numerical noise, and increases in amplitude with time until it reaches the limit cycle (full-amplitude regime). Then, normally, pulsation becomes fairly regular. An example of this behaviour is displayed in Fig.~\ref{fig:nlSTS_example1}, showing the time-dependence of several properties computed from the $\log(L/\Lsun)=3.85$ model of the sequence with $M=1.6\,\Msun$, $\aml=2.0$ and $\anu=10^{-4}$. The model reaches full amplitude after about 10 yr, when kinetic energy stops increasing and settles to a more or less constant value (except for the fluctuation over individual pulsation cycles).

For the first few years, the model in Fig.~\ref{fig:nlSTS_example1} is dominated by 1OM pulsation, in agreement with linear predictions. However, the FM soon becomes unstable giving rise to a multiperiodic behaviour seen as a modulation of the light curve around $15\,{\rm yr}\lesssim t\lesssim 30\,{\rm yr}$. At $t\gtrsim30$ yr the FM is dominant, in contrast with linear results that predict it to be stable. In other words, in this and similar models where the linear results predict stability or even weak instability, dominant FM pulsation (DFMP) in nonlinear models appears earlier (at smaller radii and lower luminosities) than in linear models. This is a common feature in our model sequences (except at the largest masses, where the FM is almost always dominant, see Sect.~\ref{ssec:NonlinearPeriodsAmplitudes}).

In general, time series can be roughly divided in two phases, i.e. linear growth and full-amplitude pulsation. When pulsation involves the FM, models go through an additional ``relaxation phase'' before attaining full amplitude \citep[cf.][their fig.~7]{Lebzelter_Wood_2005}.

This is associated with a readjustment of the envelope structure, as discussed in Sect.~\ref{ssec:EffectsOfLargeAmplitudePulsationOnStellarStructure}, and corresponds to $35\,{\rm yr}\lesssim t\lesssim 45\,{\rm yr}$ in Fig.~\ref{fig:nlSTS_example1}, where it is characterized by a ``bump'' in the trend of kinetic energy. During the relaxation phase, periods and amplitudes are systematically larger than at full amplitude.

We emphasize that the stages of amplitude growth and relaxation are not expected to be observed. These features result from the nonlinear series being initiated from a model in perfect hydrostatic equilibrium, which is not the case of pulsation in real AGB stars.

In total, we computed about 11500 nonlinear time series, corresponding to roughly 120 days of cumulative computer time using 3.6 Ghz CPUs. 
About 65 per cent (7439) of the time series were found to be stable against nonlinear pulsation, and were discarded in the present study. Note, however, that test calculation suggest that at least some of such stable models can actually pulsate and reach limit amplitude if a small velocity perturbation is applied to the initial model, rather than having pulsation grow from numerical noise. In other words, for a given luminosity sequence (i.e., for fixed stellar and model parameters), instability could emerge earlier (at lower $L$), provided the perturbation is large enough. The results presented here should be considered a ``pessimistic'' view, i.e. the largest possible luminosity at which pulsational instability is maintained.

While the perturbation approach would probably be more realistic, we chose not to follow it. Indeed, we are not especially interested in the emergence of instability (the transition from a static configuration to a pulsating one), but rather in the transition towards dominant fundamental mode pulsation, which is usually preceded by dominant pulsation in the first overtone mode.

Finally, we point out that the fact that the present models predict a range of parameters within which pulsation is stable does not mean that oscillations are absent, as they might exist by means of a different excitation mechanism. In particular, there is increasing evidence that, at luminosities typical of the upper red giant branch, LPVs undergo oscillations due to stochastic driving, similar to what happens in solar-like oscillators \citep[e.g.][]{Mosser_etal_2013}.

\subsection{Processing of time series}
\label{ssec:ProcessingOfTimeSeries}

We examine visually all time series to identify and discard stable models. For each time series, we also identified the full-amplitude portion, to which we restrict our analysis. We then estimate the mean value and amplitude of variation of relevant global properties (luminosity, surface temperature, radius) and compute their Fourier power spectrum, whose main peak we identify as the dominant period. The radial order of the dominant mode is then assessed by comparison with linear results. Pulsation in nonlinear models is never perfectly sinusoidal, hence all time series display several harmonic signals at frequencies that are integer multiples of the dominant one. Moreover, models are found along all luminosity sequences showing additional peaks corresponding to one or two pulsation modes other than the dominant. In general, this multiperiodicity is most evident when the dominant mode is about to shift.

After processing all time series, we perform a quality check to assess the reliability of results. The main issue we encounter concerns time series that are truncated early due to convergence problems. This is problematic when the time series is interrupted before attaining full amplitude, or if it covers too few full-amplitude cycles. This problem is most common in bright models dominated by FM pulsation, and is partially alleviated by retaining part of the relaxation phase of the time series in the processing step. This conservative approach is driven both by the intent of extending as much as possible the time series, and by the difficulty of distinguishing between relaxation and full-amplitude phase. The downside of this approach is that the relaxation phase is predominant in some models, hence the derived period and amplitude will overestimate the full-amplitude values.

During this quality assessment step we identified a subset of time series that we flagged as unreliable. Some of these time series are truncated well before attaining full amplitude, without any clear sign of periodicity. Other cases are more subtle as they display some degree of periodicity, but the estimated period is inconsistent with neighbouring time series, with similar mean luminosity, along model sequences. Usually, this is either due to the relaxation phase being predominant, leading to an overestimate of the pulsation period, or to some erratic behaviour that makes regular variability less evident. An example of the latter behaviour is displayed in Fig.~\ref{fig:nlSTS_example2}, showing a time series in which the outer mass zones are pushed to large distances from the nominal surface and follow long ballistic trajectories before falling back.

Overall, such flagged models represent only about 6 per cent of our set, but are in most cases found at large luminosities. They are therefore retained as indicators of upper limits of period and amplitude in order to better characterize this regime.

\begin{figure*}
    \includegraphics[width=.99\textwidth]{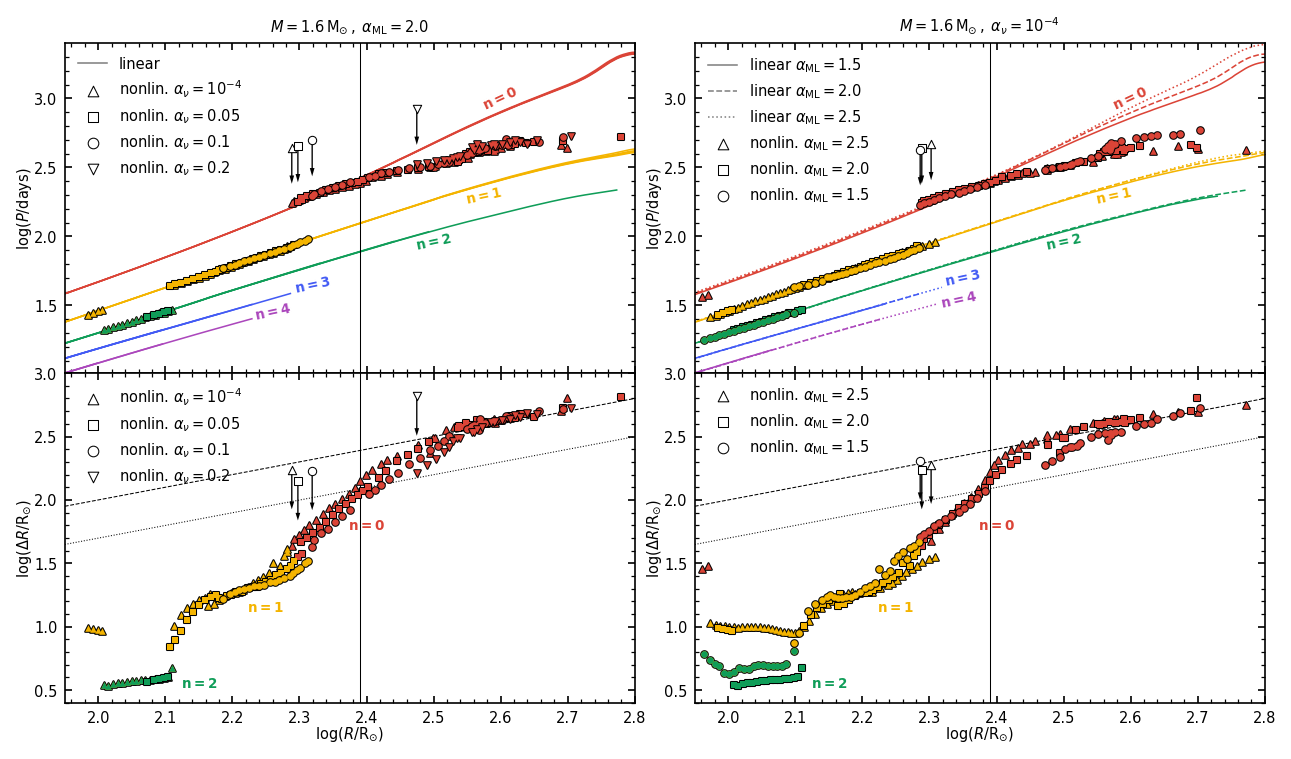}
    \caption{
    Panels on the left show the dependence on surface radius of nonlinear periods (top) and peak-to-peak amplitude of surface displacement $\Delta R$ (bottom), as a function of varying $\anu$ (different symbols), while $\aml$ is fixed. The same quantities are shown in the panels on the right, except $\anu$ is fixed and symbols indicate different values of $\aml$. Linear periods are shown as solid lines in the top panel for reference. Radial orders of pulsation are colour-coded as in Fig.~\ref{fig:linear_anueffects_example}. Arrows mark the radius at which fundamental mode pulsation becomes dominant, depending on the value of $\anu$ (left panels) or $\aml$ (right panels). Vertical lines indicate approximately the radius at which nonlinear fundamental mode periods deviate from linear predictions. Dashed and dotted lines in the bottom panel correspond to $\Delta R=R$ and $\Delta R=0.5\,R$, respectively.
    }
    \label{fig:nonlinearPAmp_var_aML_anu}
\end{figure*}

\begin{figure*}
    \includegraphics[width=.99\textwidth]{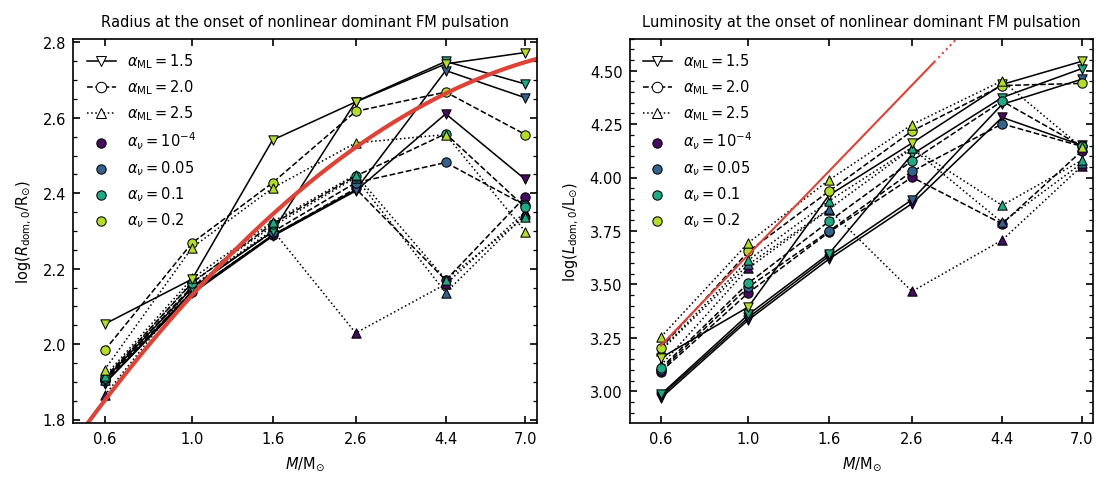}
    \caption{
    Radius $R_{\rm dom,0}$ (left panel) and luminosity $L_{\rm dom,0}$ (right panel) at which the fundamental mode becomes dominant in the present nonlinear calculations, as a function of mass. Different symbols are used to indicate distinct values of the mixing length parameter (downward pointing triangles: $\aml=1.5$; circles: $\aml=2.0$; upward pointing triangles: $\aml=2.5$), while the value of the turbulent viscosity parameter $\anu$ is colour-coded. The solid red line indicates the approximate relation we adopted to describe long-period variability in synthetic stellar population models (see text). Note that the scale is logarithmic along the horizontal axis. For comparison, the red line in the right panel indicate the onset of dominant fundamental mode pulsation according to linear predictions, obtained from Eq.~10 of \citet{Trabucchi_etal_2017} (strictly valid only for $M\lesssim3\,\Msun$, and corresponding to $\anu=0$).
    }
    \label{fig:nlP0_onset}
\end{figure*}

\begin{figure*}
    \includegraphics[width=\textwidth]{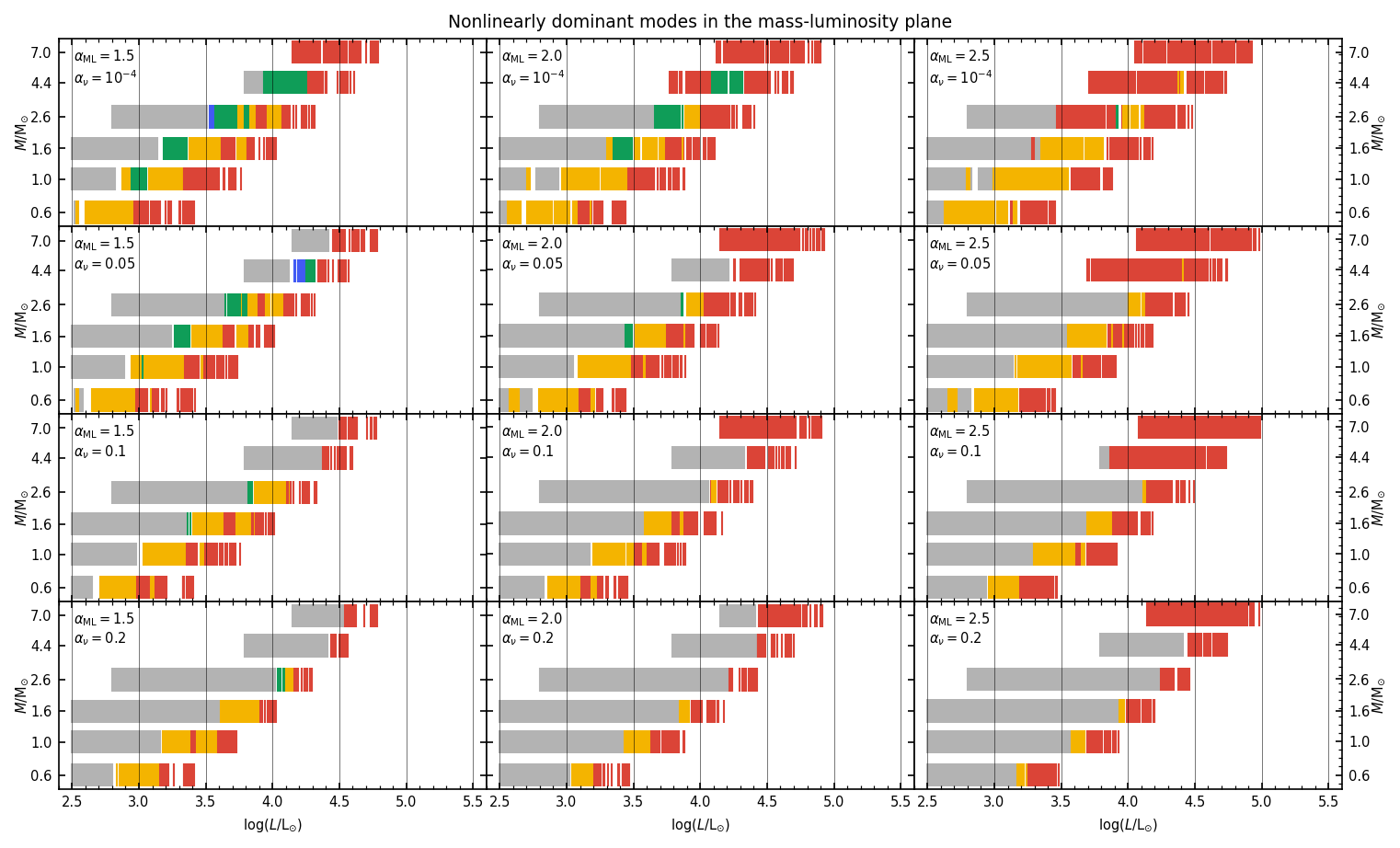}
    \caption{
    Similar to Fig.~\ref{fig:linearDomPatches_ML}, but for the results of nonlinear models.
    }
    \label{fig:nonlinearDomPatches_ML_noFlag_stable}
\end{figure*}

\begin{figure*}
    \includegraphics[width=\textwidth]{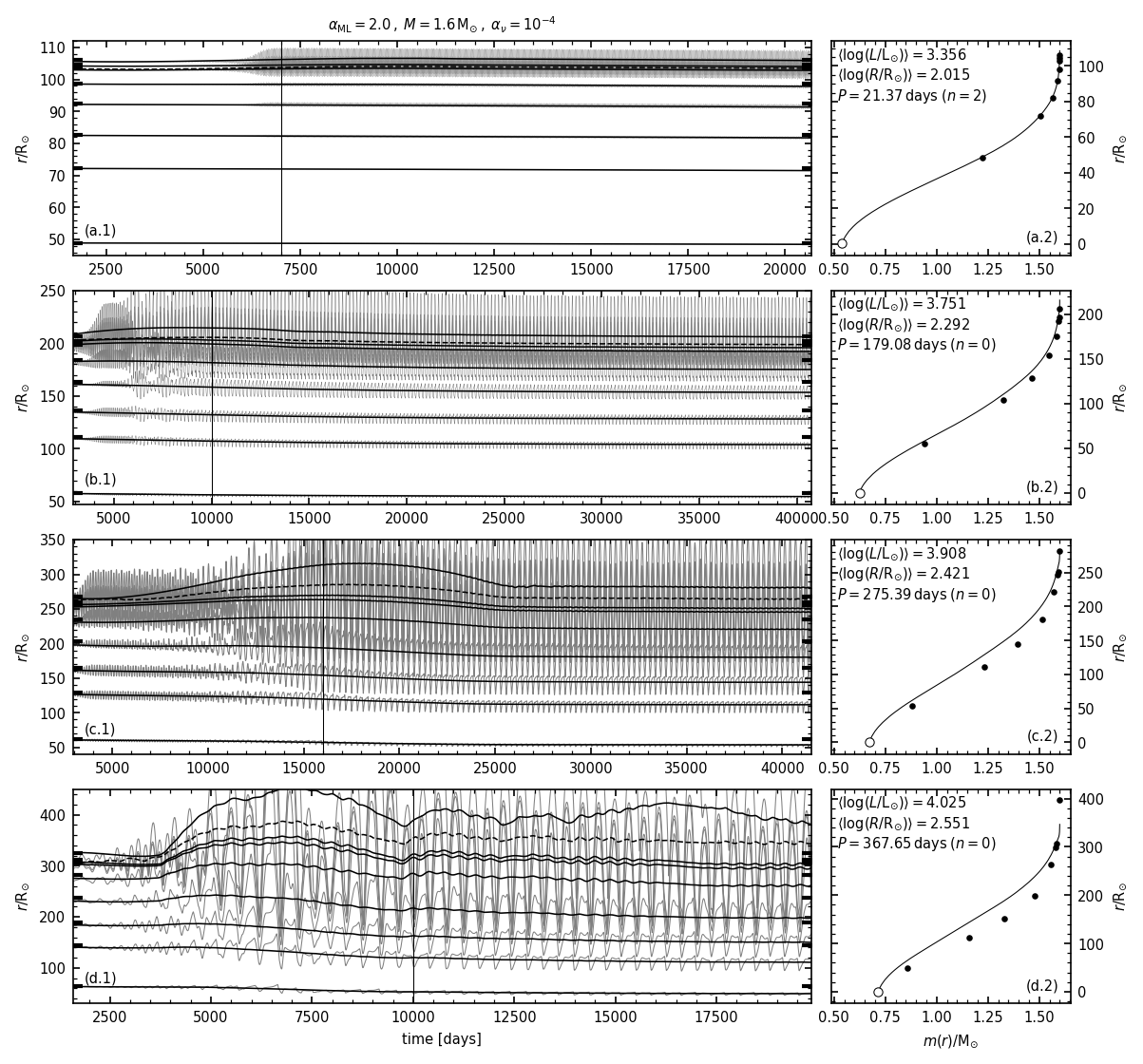}
    \caption{
    Pulsation in the envelope interior of four models at different stages along the sequence with $M=1.6\,\Msun$, $\aml=2.0$, $\anu=10^{-4}$. Rows from top to bottom show nonlinear results corresponding to static models with $\log(L/\Lsun)=3.36$, 3.77, 3.92, 4.03. Panels in the left column show the radial displacement as a function of time of a few selected mass zones within the nonlinear models. Light solid lines include each pulsation cycle, while pulsation has been smoothed out in the dark solid lines, the dashed line showing the position of the optical surface. Large ticks along the vertical axes mark the radii of the same mass zones in the corresponding hydrostatic model. Vertical lines indicate the time at which the model has approximately reached (or is relaxing to) full-amplitude pulsation. Panels in the right column show the time-averaged interior mass distribution of nonlinear models (circles) in comparison with the mass profile of the corresponding static envelope models. Empty circles mark the edge of the rigid core. Mean values of luminosity and radius are indicated in the panels on the right.
    }
    \label{fig:nlSTS_adjust}
\end{figure*}

\section{Results and discussion}
\label{sec:Results}

\subsection{Nonlinear periods and amplitudes}
\label{ssec:NonlinearPeriodsAmplitudes}

The general trend of nonlinear periods $P$ and amplitudes $\Delta R$ as a function of mean stellar radius is illustrated in Fig.~\ref{fig:nonlinearPAmp_var_aML_anu}, which displays the values determined for the $M=1.6\,\Msun$ sequences. The stellar radius $R$ is defined by the location where the Rosseland mean optical depth is $\tau_{\rm R}=2/3$, and $\Delta R$ is its total variation in the limit cycle model. As in the linear case, periods depend only weakly (and only at large radii) upon $\aml$, while they are essentially insensitive to the value of $\anu$. The only exception is the low-luminosity regime of the most massive models ($M\geq4.4\,\Msun$), in which the FM period shows a slight dependence upon $\anu$.

The most striking feature, however, is the stark deviation from linear predictions at $\log(R/\Rsun)\gtrsim2.4$, beyond which the fundamental mode PR relation becomes significantly less steep. In contrast, linear and nonlinear FM periods are in good agreement at smaller radii. In the present nonlinear models, the period of dominant overtone mode pulsation is always found to be in excellent agreement with linear results.

The deviation of the nonlinear FM period from the linear FM period is due to the rearrangement of the stellar envelope structure that occurs for large-amplitude FM pulsation \citep[][, see Sect.~\ref{ssec:EffectsOfLargeAmplitudePulsationOnStellarStructure}]{YaAri_Tuchman_1996,Lebzelter_Wood_2005,Kamath_etal_2010}. As seen in Fig.~\ref{fig:nonlinearPAmp_var_aML_anu}, the amplitude $\Delta R$ of surface displacement increases monotonically with stellar radius and the nonlinear periods diverge from linear predictions when $\Delta R\sim0.5\,R$. The increase of amplitude with radius slows down as it reaches $\Delta R\sim R$, that appears to be a limiting value. This will be discussed in more detail in Sect.~\ref{sssec:ChangeOfSlope}.

The effect of varying model parameters $\aml$ and $\anu$ upon amplitude is not obvious, mostly because the values of $\Delta R$ displayed in the bottom panels of Fig.~\ref{fig:nonlinearPAmp_var_aML_anu} include pulsation from all active modes, and not only the dominant. In general, amplitudes become smaller when $\anu$ is increased, i.e. when turbulent viscous dissipation is larger. However, this seems to be the case only for the FM, while overtone modes appear to be largely unaffected. A rough quantitative estimate suggests that $\Delta R$ is reduced by 30-40 per cent when $\anu$ is increased from 0 to 0.2, which is similar to the effect obtained by decreasing $\aml$ from 2.5 to 1.5.

An additional effect of varying mixing length and turbulent viscosity is to shift the stability regime of all modes, in particular the value of radius $R_{\rm dom,0}$ (or luminosity $L_{\rm dom,0}$) at which the FM becomes dominant. This radius  is indicated by arrows in Fig.~\ref{fig:nonlinearPAmp_var_aML_anu}, and is shifted towards larger values by increasing $\anu$. In contrast, $R_{\rm dom,0}$  is little affected by changing $\aml$, at least at small masses. Fig.~\ref{fig:nlP0_onset} gives a more general picture, showing the dependence of $R_{\rm dom,0}$ and $L_{\rm dom,0}$ upon mass, $\anu$, and $\aml$, and a comparison with the linear prescription for the onset of DMFP from \citetalias{Trabucchi_etal_2019}. At the larger masses, the pattern is complicated by the suppression of overtone modes, which favours a much earlier onset of DFMP.

It is instructive to examine the regions where nonlinear pulsation is dominant in the mass-luminosity plane, which is displayed in Fig.~\ref{fig:nonlinearDomPatches_ML_noFlag_stable} (the nonlinear equivalent of Fig.~\ref{fig:linearDomPatches_ML}). Note that model sequences at each mass are less extended in luminosity than in the case of linear calculations, due to the difficulty of converging nonlinear time series at large luminosities. These issues and the occurrence of flagged models result in the white stripes visible in Fig.~\ref{fig:linearDomPatches_ML}. With respect to Fig.~\ref{fig:linearDomPatches_ML}, these stripes are further emphasized by the fact that the mean luminosity of nonlinear time series is slightly different than that of the initial static model, resulting in an uneven luminosity sampling. White gaps should thus not be interpreted as regions of pulsational stability.

There are two main differences with respect to Fig.~\ref{fig:linearDomPatches_ML}.
Firstly, the low-luminosity portion of each sequence is much less populated (grey patches). This may be because the nonlinear models were not run long enough for pulsation to grow from noise. Recalling the remarks made in Sect.~\ref{ssec:ComputationAndGeneralFeatures} that a small imposed perturbation can lead to a continuing pulsation in seemingly stable models, the grey patches in Fig.~\ref{fig:nonlinearDomPatches_ML_noFlag_stable} should not immediately be interpreted as evidence for pulsational stability. Nonetheless, the results in Fig.~\ref{fig:linearDomPatches_ML} do suggest that nonlinear pulsation hardly occurs in overtones higher than the second, and that the fundamental mode does not show two dominance regimes in the nonlinear case. Secondly, and most importantly, the transition to DFMP occurs at lower luminosities than in the linear case, and the same is true for overtone modes (when a comparison is possible). This is especially evident at small values of $\aml$ and $\anu$, with DFMP arising as much as $\sim0.3$ dex earlier in luminosity, an amount that is reduce to $\sim0.05$ dex when $\aml=2.5$ and $\anu=0.2$.

\subsection{Effects of large-amplitude pulsation on stellar structure}
\label{ssec:EffectsOfLargeAmplitudePulsationOnStellarStructure}

As discussed above, the onset of large-amplitude, FM pulsation causes the nonlinear PR relation to deviate from linear predictions. To explain this, we examine how the average envelope structure of nonlinear models is affected by pulsation. We consider four time series at different stages along the luminosity sequence with $M=1.6\,\Msun$, $\aml=2.0$, and $\anu=10^{-4}$. The interior pulsation of these models is displayed in Fig.~\ref{fig:nlSTS_adjust} in terms of the radial motion of selected mass zones. Panel (a.1) displays a relatively small-radius model ($\simeq103\,\Rsun$) whose pulsation is dominated by the 2OM. Pulsation has low amplitude and is limited to the outermost layers. In comparison, pulsation is significant throughout the outer half of the envelope of the $\sim196\,\Rsun$ model displayed in panel (b.1). This model is dominated by FM pulsation, but is still in the linear regime. While amplitude in its interior is much smaller than at the surface, it is large enough to cause a slight readjustment of the envelope structure, that shows a tendency to contract (except for the outermost layers). The structural readjustment can be better appreciated in terms of the mass profile $m(r)$, as displayed in panels (a.2) and (b.2) where the static structure (solid line) is compared with the time-averaged structure of the nonlinear model (circles).

The model shown in panels (c.1) and (c.2) has $R\simeq264\,\Rsun$ and is well beyond the linear regime. Pulsation occurs with significant amplitude in the outer 60 per cent of the envelope by radius, causing a substantial readjustment of the envelope. In the $R\simeq356\,\Rsun$ model [panels (d.1) and (d.2)] mass zones are displaced by as much as 20-25 per cent with respect to the static structure.

We conclude that large-amplitude pulsation causes the stellar envelope to develop a steeper density stratification than what would occur if it were in hydrostatic equilibrium. In turn, this affect pulsation periods, that become shorter. This behaviour of nonlinear models was first noted by \citet{YaAri_Tuchman_1996}, and can be understood, on a qualitative level, in terms of the period-mean density relation, $P\propto\overline{\rho}^{-1/2}$ \citep[e.g.][]{Cox_TSP_1980}, even though it strictly applies only in hydrostatic equilibrium. The majority of the envelope layers that undergo pulsation have higher density with respect to the static case, and therefore pulsate with shorter period.

\begin{figure*}
    \includegraphics[width=\textwidth]{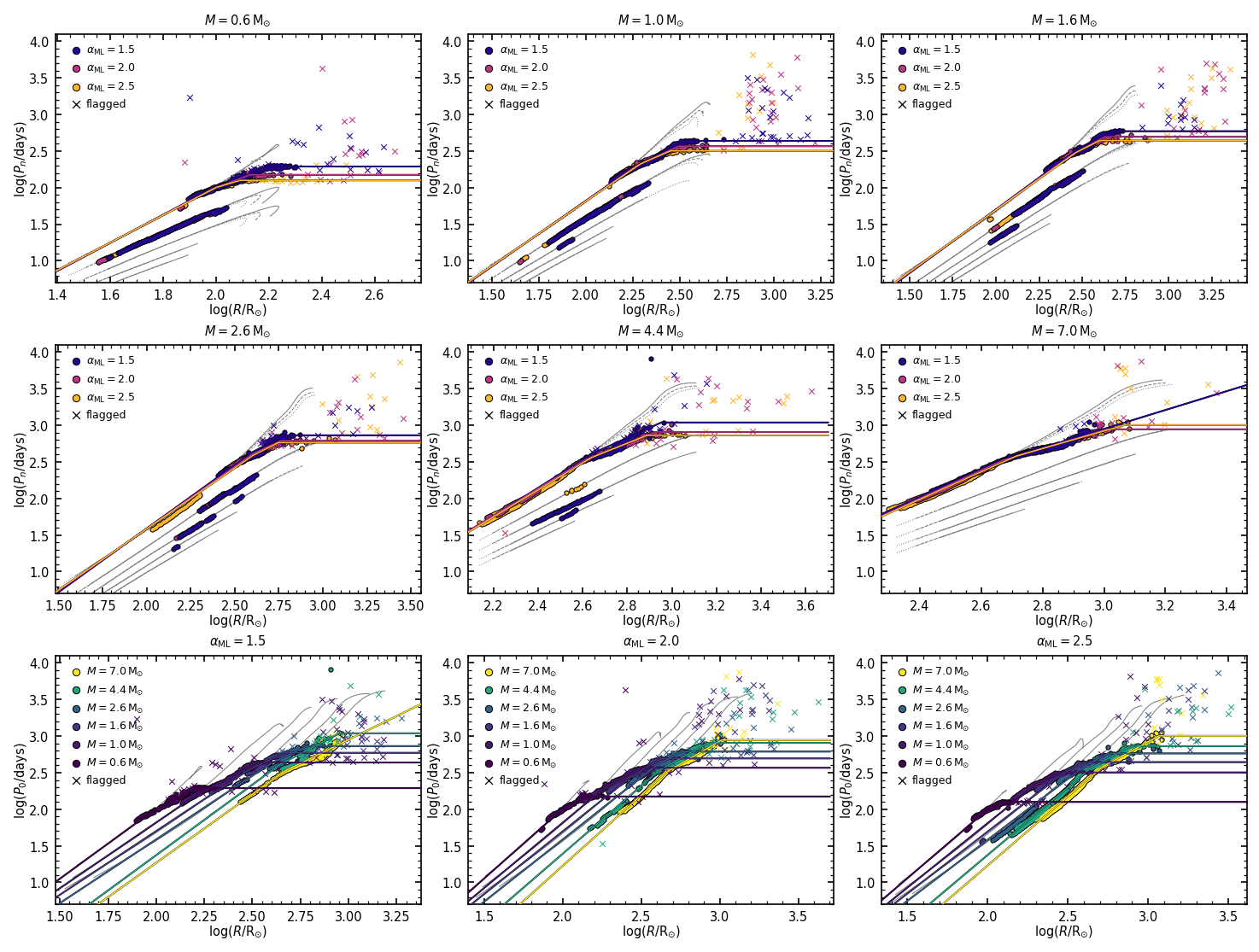}
    \caption{
    Period-radius relations from nonlinear models at fixed mass but varying $\aml$ (panels in the top and middle rows), and at fixed $\aml$ but varying mass (panels in the bottom row). Light gray lines in the background indicate predictions of linear models. No distinction is made between models with different $\anu$. Cross symbols indicate flagged models. Coloured solid lines are the best fits of Eq.~\ref{eq:nlPMR} for each combination of mass and $\aml$ (colour coding is indicated in each the panel).
    }
    \label{fig:nonlinearPMRfit_M_alpha}
\end{figure*}

\begin{figure*}
    \includegraphics[width=\textwidth]{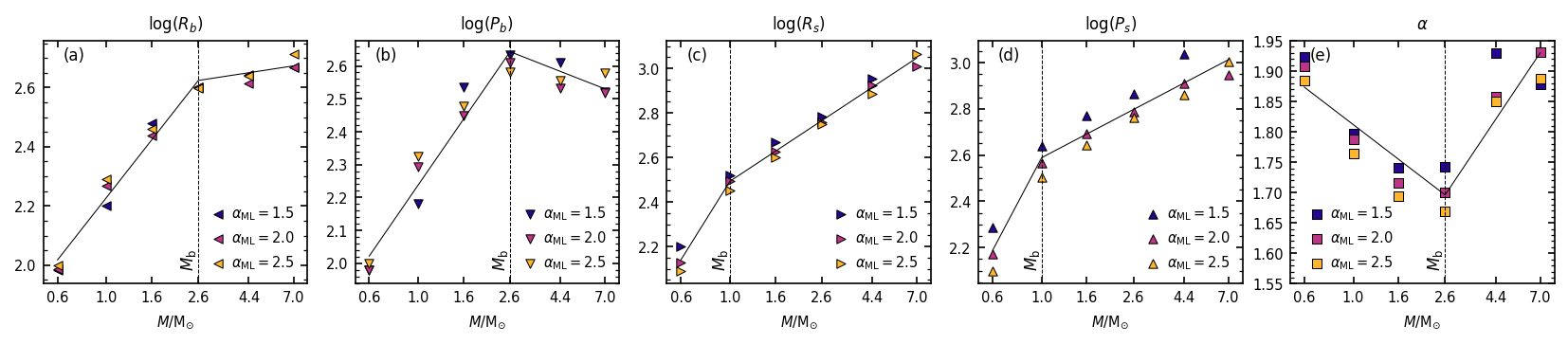}
    \caption{
    Best-fit parameters for the period-radius relation (Eq.~\ref{eq:nlPMR}) as a function of mass and mixing length parameter $\aml$ (colour-coded). Note that the scale is logarithmic along the horizontal axis.
    }
    \label{fig:fitParam_M_alpha}
\end{figure*}

\subsection{Nonlinear period-mass-radius relation of the fundamental mode}
\label{ssec:NonlinearPeriodMassRadiusRelationRelationOfTheFundamentalMode}

\subsubsection{Best-fit PMR}
\label{sssec:bestfitPMR}
One of the key objectives of this study is to deliver an easy-to-use tool to make predictions of pulsation properties as a function of global stellar parameters. A convenient way to do so is by deriving analytical approximations of the period-mass-radius (PMR) relationship followed by pulsation models. These are displayed in Fig.~\ref{fig:nonlinearPMRfit_M_alpha}. Each panel there corresponds to a fixed mass, except for panels in the bottom row in which $\aml$ is fixed instead. Linear results are also shown for reference. It is evident that linear and nonlinear overtone mode periods are essentially equal at all masses. Hence we focus on the fundamental mode, for which we derive a nonlinear PMR in two steps: (1) we examine the PR relation at fixed mass, and identify an adequate best-fit analytic function to represent it; (2) we model the dependence of the best fitting parameters upon mass.

Fig.~\ref{fig:nonlinearPMRfit_M_alpha} clearly shows the two features of the PR relation introduced in Sect.~\ref{ssec:NonlinearPeriodsAmplitudes}, i.e. the deviation from the linear trend and the tendency to approach a constant period at large radii. We model these features by adopting the following functional form:
\begin{equation}\label{eq:nlPMR}
    \log(P_0) =
    \left\{
        \begin{array}{ll}
            \log(\Pb) + \alpha\log(R/\Rb)   & \mbox{if } R<\Rb \\
            \log(\Pb) + \beta\log(R/\Rb)    & \mbox{if } \Rb\leq R<\Rs \\
            \log(\Ps)                       & \mbox{if } \Rs\leq R \,,
        \end{array}
    \right.
\end{equation}
where radii are in solar units and periods in days. In the $\log(P)$~-~$\log(R)$ plane, Eq.~\ref{eq:nlPMR} represents a broken line whose slope changes from $\alpha$ to $\beta=\log(\Pb/\Ps) / \log(\Rb/\Rs)$ at the breaking point ($\Rb$, $\Pb$), and becomes zero after the ``saturation'' point ($\Rs$, $\Ps$). In other words we describe the period-radius relation as a power law broken at two points, with a saturation at large radii (where period becomes independent of radius).

We constrain the values of the five free parameters $\Rb$, $\Pb$, $\Rs$, $\Ps$, and $\alpha$ by computing a least-square fit to the data points obtained from models, and the result is displayed by solid lines in Fig.~\ref{fig:nonlinearPMRfit_M_alpha}. This is done independently for each combination of mass and $\aml$, while no distinction is made between models computed with different $\anu$. Moreover, we exclude data points corresponding to flagged models, i.e. to time series for which the estimated period is not considered reliable (see Sect.~\ref{ssec:ProcessingOfTimeSeries}). Best-fit parameters as a function of mass and $\aml$ are shown in Fig.~\ref{fig:fitParam_M_alpha}.
Even though there is some dependence upon $\aml$, we do not model it (see Sect.~\ref{sssec:remarks}). The dependence of each of the parameters upon mass can be described once again by a broken power law. Given the coarse sampling, the breaking mass $M_{\rm b}$ cannot be determined accurately by least-square fitting, but is close to $\Mb=1.0\,\Msun$ (for $\Rs$ and $\Ps$) and $\Mb=2.6\,\Msun$ (for $\Rb$, $\Pb$, and $\alpha$). It is convenient to adopt these values. This leaves us with two free parameters to be determined in the adopted the functional form:
\begin{equation}\label{eq:nlMdep}
    K =
    \left\{
        \begin{array}{ll}
            \log(\kb) + \gamone\log(M/\Mb)   & \mbox{if } M<\Mb \\
            \log(\kb) + \gamtwo\log(M/\Mb)    & \mbox{if } \Mb\leq M \,,
        \end{array}
    \right.
\end{equation}
where $K=\log(\Rb)$, $\log(\Pb)$, $\log(\Rs)$, $\log(\Ps)$, or $\alpha$. The hyper-parameters $\kb$ and $\Mb$ represent coordinates of breaking points past which the slope changes from $\gamone$ to $\gamtwo$. The best-fit curves are displayed in Fig.~\ref{fig:fitParam_M_alpha}, while the best-fit values of the twenty hyper-parameters derived this way are summarized in Table~\ref{tab:hyperparameters}.

\begin{figure}
    \includegraphics[width=\columnwidth]{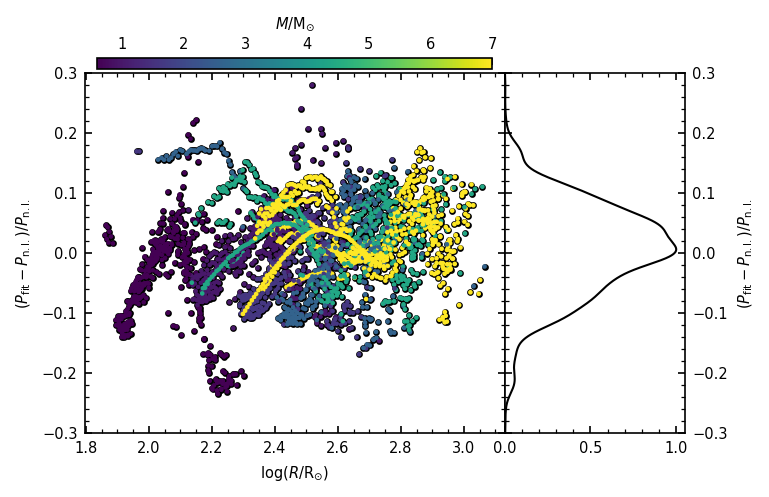}
    \caption{
    Relative error made when computing the fundamental mode period from Eq.~\ref{eq:nlPMR} with respect to the actual values determined from nonlinear time series. Predictions are within a 10 per cent error in more than 80 per cent of the cases.
    }
    \label{fig:fitErrors}
\end{figure}

To test our best-fit PMR, we use Eqs.~\ref{eq:nlPMR} and~\ref{eq:nlMdep} to estimate periods from mass and radius for the original data set (once again, excluding flagged models). The resulting periods are within 10 per cent of the original value in more than 80 per cent of cases, as displayed in Fig.~\ref{fig:fitErrors}. The relative error is not found to correlate significantly with any relevant stellar or model parameter.

\begin{table}
    \centering
    \caption{
        Best-fit values of the hyper-parameters of Eq.~\ref{eq:nlMdep}, which describe the dependence upon mass of the parameters of Eq.~\ref{eq:nlPMR}.
        }\label{tab:hyperparameters}
    \begin{tabular}{c|c|c|c|c}
        Parameter & $\kb$ & $\Mb$ & $\gamone$ & $\gamtwo$ \\
        \hline
        $\log(\Rb)$ & $421\,\Rsun$      & $2.6\,\Msun$ &  0.952 &  0.114 \\
        $\log(\Pb)$ & $440\,{\rm days}$ & $2.6\,\Msun$ &  0.976 & -0.264 \\
        $\log(\Rs)$ & $311\,\Rsun$      & $1.0\,\Msun$ &  1.590 &  0.654 \\
        $\log(\Ps)$ & $388\,{\rm days}$ & $1.0\,\Msun$ &  1.808 &  0.502 \\
        $\alpha$    & $49.7$            & $2.6\,\Msun$ & -0.279 &  0.544
    \end{tabular}
\end{table}

\subsubsection{The change of slope}
\label{sssec:ChangeOfSlope}
An unfortunate consequence of the change in slope of the PR relation is that the nonlinear PMR relation cannot be uniquely inverted to estimate mass and radius from a given period. This can be appreciated in the bottom-row panels of Fig.~\ref{fig:nonlinearPMRfit_M_alpha}, where it is evident that the period-radius relations are seen to cross. This makes it somewhat difficult to constrain the global parameters of LPVs from their observed periods. It is reasonable to think that this degeneracy could be lifted by using information on pulsation amplitudes, a detailed understanding of which is thus desirable.

\begin{figure}
    \includegraphics[width=\columnwidth]{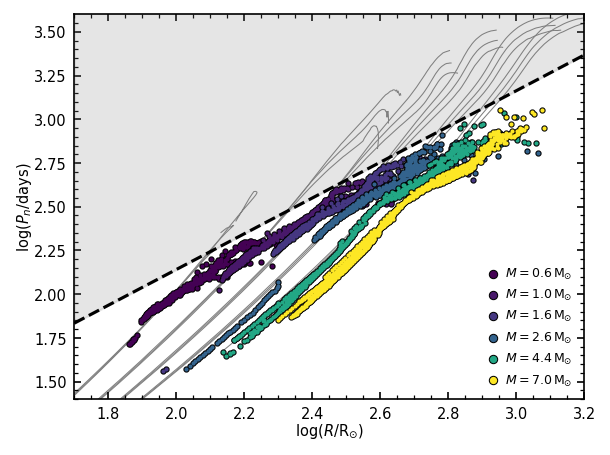}
    \caption{
    Region of avoidance of nonlinear pulsation (shaded area) in the PR diagram. Circles have the same meaning as in the bottom-row panels of Fig.~\ref{fig:nonlinearPMRfit_M_alpha}. Flagged models are not shown. The dashed line is defined by Eq.~\ref{eq:DeathLine}.
    }
    \label{fig:DeathLine}
\end{figure}

\begin{figure*}
    \includegraphics[width=\textwidth]{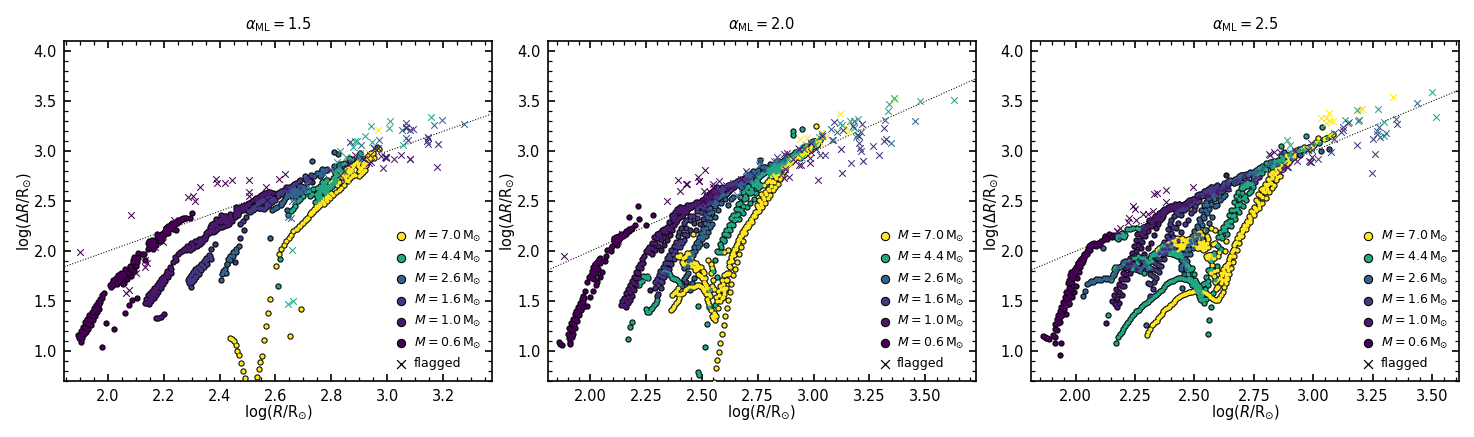}
    \caption{
    Similar to the bottom row of Fig.~\ref{fig:nonlinearPMRfit_M_alpha}, but showing the amplitude $\Delta R$ of displacement of the stellar surface during pulsation as a function of mean stellar radius. Dashed lines correspond to $\Delta R = R$. Each panel correspond to a different value of $\aml$, while mass is colour-coded and no distinction is made for different values of $\anu$.
    }
    \label{fig:nonlinearDR_aML}
\end{figure*}

On the other hand, the bending and flattening of the PR relation results in the existence of a region of the PR diagram where nonlinear pulsation is never found, regardless of mass, in contrast with linear results. In other words nonlinear pulsation periods cannot grow beyond a ``death line'' (the dashed line in Fig.~\ref{fig:DeathLine}) approximately described by
\begin{equation}\label{eq:DeathLine}
    \log(P_{\rm 0,max}) = 0.096 + 1.022 \log(R) \,,
\end{equation}
which represents the maximum FM period at a given radius. Given the observed period of an LPV known to pulsate in the FM, Eq.~\ref{eq:DeathLine} can be used to estimate a lower limit to its surface radius.

It is worth noting that the amplitude $\Delta R$ of surface displacement shows a somewhat similar behaviour to that of periods. As noted in Sect.~\ref{ssec:NonlinearPeriodsAmplitudes}, its growth is limited to values smaller than or approximately equal to the current surface radius. Once this limit is reached, the rate of amplitude increase is set by $\Delta R\simeq R$. This property is shared by all our sequences of nonlinear models, and appears to be independent of model parameters $\aml$ and $\anu$, as displayed in Fig.~\ref{fig:nonlinearDR_aML}. The existence of some dissipative process preventing an arbitrarily large growth of pulsation amplitude would not be surprising, however the reason behind the behaviour displayed by models, and in particular why amplitude should be limited to the value of the surface radius, is not clear, and will not be further investigated here.

\subsubsection{Remarks}
\label{sssec:remarks}
Some assumptions have been made in deriving the analytic PMR that deserve discussion. Having excluded flagged models from the fit makes questionable the choice of a flattening PR relation. At large masses, in particular, the large-radii regime is poorly constrained, and the saturation point is essentially determined by the longest non-flagged period in the data set. The sequences with $M=7.0\,\Msun$ and $\aml=1.5$ represent an extreme case in which the saturation parameters cannot be constrained at all. Nonetheless, there is evidence in support of our choice. Let us consider, for instance, the $M=1.0\,\Msun$ sequence (central panel in the top row of Fig.~\ref{fig:nonlinearPMRfit_M_alpha}). Most of the flagged models are distributed irregularly in the top-right corner of the diagram, but a significant fraction are well consistent with the best-fit curve, even though they were excluded from its derivation. The situation is similar for other values of mass, except the largest ones. One could doubt that the parameters $\Rb$ and $\Pb$ obtained for $M\geq4.4\,\Msun$ are realistic, given the scarcity of data points to fit, but panels (c) and (d) of Fig.~\ref{fig:fitParam_M_alpha} suggest otherwise. Indeed, the best-fit values at large masses are well consistent with the trend observed at small masses, where these parameters are better constrained. A dedicated study of nonlinear pulsation in massive AGB models is desirable to better understand this.

It is also worth spending a few words about the role of $\aml$ in the PMR relation, which we completely neglected. This approximation is rather good and justified for all parameters of Eq.~\ref{eq:nlPMR}, except the saturation period $\Pb$. This is immediately clear by looking at the top-left panel of Fig.~\ref{fig:nonlinearPMRfit_M_alpha}: changing $\aml$ from 1.5 to 2.5 leads to a 60 per cent difference in $\Pb$ for the $1.0\,\Msun$ models. The difference is somewhat smaller at larger masses [see also panel (d) of Fig.~\ref{fig:fitParam_M_alpha}], but still significant. The main problem is that the value of $\aml$ is meaningful only in the context of the codes used in this study, hence its inclusion in the PMR relation would only lead to confusion. In principle, one could examine the dependence of the PMR on some other stellar parameter related with $\aml$, such as effective temperature. We explored this possibility and found no description simple enough to justify its inclusion in the best-fit PMR. Disregarding the dependence upon $\aml$ is therefore justified in terms of best trade-off between accuracy and simpleness of the analytic PMR relation.

\subsection{Effect of chemical composition}
\label{ssec:EffectOfChemicalComposition}

The effects on linear pulsation of independent changes in metallicity and $\co$, at fixed mass and radius, have been examined in \citetalias{Trabucchi_etal_2019}. It was shown that, as far as overtone modes are concerned, increasing the metallicity of an O-rich model from $Z=0.006$ to $Z=0.017$ causes linear periods to shorten by less than 1 per cent. In contrast, the same change can decrease the linear FM period by up to $\sim10$ per cent. A similar change can be achieved by varying $\co$ from $\sim0.55$ ($\simeq\co_{\odot}$) to 3. Note, however, that the latter result was obtained in \citetalias[][]{Trabucchi_etal_2019} (as well as in the present work) by increasing the mass fraction $\xc$ of carbon at the expenses of helium, while keeping fixed the abundance of oxygen $\xo$. While this is qualitatively consistent with evolutionary chemical enrichment, it also means that we cannot completely disentangle the effects of pure $\co$ variations from those associated with increased metallicity.

In terms of stability, increasing the metallicity in the linear models was found to lead to a decrease in growth rates of both the FM and 1OM, causing a delay in the onset of dominant pulsation in these modes, while leaving higher overtones almost unaffected. Growth rates decrease with increasing $\co$ (except they have a maximum around $\co\simeq1$), therefore, in general, the onset of pulsation is delayed in C-rich models with respect to O-rich ones.

\begin{figure}
    \includegraphics[width=\columnwidth]{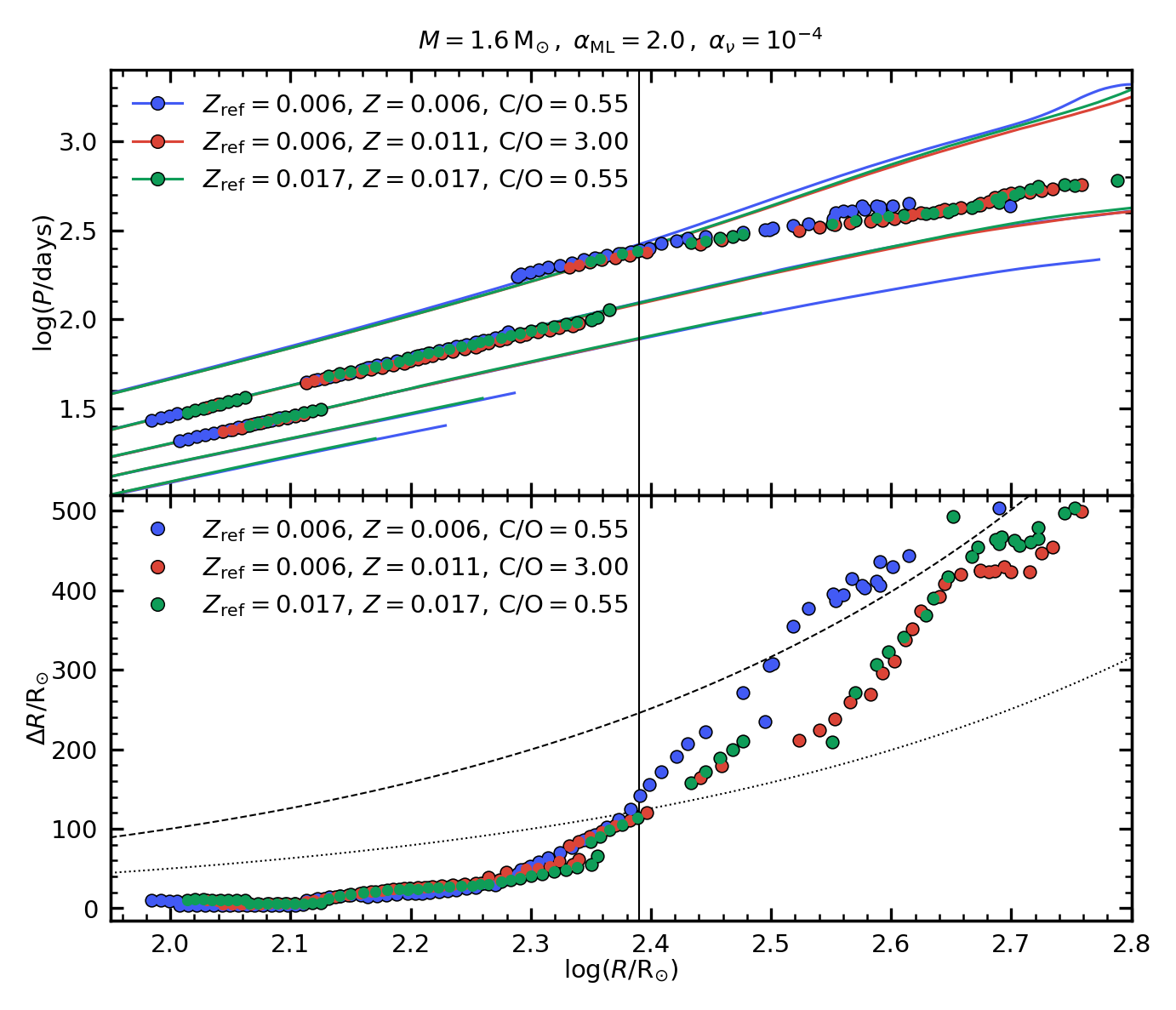}
    \caption{
    Periods (top panel) and amplitudes (bottom panel) of an O-rich model sequence (blue circles) and for the corresponding C-rich (red circles) and solar-metallicity (green circles) test sequences. All sequences are computed with $M=1.6\,\Msun$, $\aml=2.0$, and $10^{-4}$. Linear periods are shown in the top panel for reference, using the same colour code. As in Fig.~\ref{fig:nonlinearPAmp_var_aML_anu}, vertical solid lines mark approximately the radius at which nonlinear fundamental mode periods deviate from linear predictions, while the dashed and dotted lines in the bottom panel represent $\Delta R=R$ and $\Delta R=0.5\,R$, respectively.}
    \label{fig:nonlinPamp_Crich_Zsun}
\end{figure}

Given the substantial chemical evolution that AGB stars are subject to, and the variety of astrophysical environments they populate, a systematic analysis of the impact of varying chemical composition upon nonlinear pulsation is highly desirable, but is beyond the scope of this work. Here, we probe possible such effects by considering only two test cases corresponding, respectively, to C-rich composition and solar metallicity. We compute luminosity sequences for these two cases setting the other parameters to $M=1.6\,\Msun$, $\aml=2.0$, and $\anu=10^{-4}$. The results are displayed in Fig.~\ref{fig:nonlinPamp_Crich_Zsun}.

Qualitatively, we find the same trends identified in the linear case. In the regime where the FM is dominant, C-rich and solar-metallicity models have nonlinear periods about $10-15$ per cent shorter than the reference models (O-rich and with $Z=0.006$). Moreover, the onset of DFMP is delayed (occurs at larger radii and higher luminosities), and has smaller amplitude at a given radius, both features being consistent with the stability predictions from linear models. Another effect of composition variations is the delay of the amplitude ``saturation'', which occurs at $R\gtrsim350\,\Rsun$ in the reference O-rich sequence, and at $R\gtrsim450\,\Rsun$ in the C-rich and solar-metallicity sequences. According to the discussion of Sect.~\ref{ssec:NonlinearPeriodMassRadiusRelationRelationOfTheFundamentalMode}, this suggests that the change of slope of the PR relation could also be delayed. However, our test cases do not provide enough evidence to support this possibility. Further investigation is needed to understand this and to make predictions concerning possible observable effects.

\section{Comparison with observations}
\label{sec:ComparisonWithObservations}

\subsection{Simulated variability}
In order to compare our results with observations of variability in LPVs, we follow the same general procedure described in \citet{Trabucchi_etal_2017}. As reference observations we adopt data from the third phase of the Optical Gravitational Lensing Experiment \citep{Udalski_etal_1992}, namely the OGLE-III Catalog of LPVs in the Magellanic Clouds \citep{Soszynski_etal_2009_LMC,Soszynski_etal_2011_SMC}, which sources we cross-match with near-infrared photometry from the Two Micron All Sky Survey \citep[2MASS, ][]{Cutri_etal_2003,Skrutskie_etal_2006}. We complement this data set with variability information and photometry from the second data release of the \gaia\ mission \citep[\gaia\ DR2,][]{GaiaCollaboration_2018}, namely the catalog of LPVs by \citet{Mowlavi_etal_2018}. From the latter, we select stars belonging to the LMC and SMC using the criteria of \citet{Mowlavi_etal_2019} (their table~1).

We use the \trilegal\ code \citep{Girardi_etal_2005} to produce synthetic models of the AGB population of the Magellanic Clouds. Stellar isochrones used in \trilegal\ \citep{Marigo_etal_2017} are obtained from \parsec\ evolutionary tracks \citep{Bressan_etal_2012}, and from TP-AGB tracks computed with the \colibri\ code \citep{Marigo_etal_2013}. In particular, we use the TP-AGB tracks from the S\_35 set by \citet{Pastorelli_etal_2019} and from the S\_37 set by \citet{Pastorelli_etal_2020}. We refer to these works for further details concerning the setup of the simulations.

We filter the simulation to exclude evolutionary stages other than the AGB, except we retain simulated stars experiencing late core-He burning (CHeB), representative of red supergiants (RSGs). This choice is motivated by the presence of RSG semi-regular variables in the \gaia\ DR2 LPV that display a similar behaviour to AGB LPVs, and for which a comparison between models and observations is interesting. Strictly speaking, pulsation models presented here are not appropriate to describe pulsation in RSGs due to the chosen core mass-luminosity relation that does not reflect the properties of CHeB stars. Keeping that in mind, we assume that the core mass and size have little effect on pulsation properties.

We further restrict the simulation to O-rich stars only, i.e. with $\co<1$. Observational data are filtered accordingly following the approach of \citet{Lebzelter_etal_2018} based on the \gaia-2MASS diagram. We compute linear periods and growth rates for each simulated star by interpolating in the grid of models from \citetalias{Trabucchi_etal_2019}. Variability properties are computed this way for all pulsation modes from the fundamental to the fourth overtone. However, we focus our analysis on FM pulsation, as nonlinear overtone periods are entirely compatible with linear predictions and in agreement with observations \citep{Trabucchi_etal_2017}. Finally, we estimate nonlinear FM periods for simulated stars using by applying the best-fit PMR derived in Sect.~\ref{sssec:bestfitPMR}.

Pulsation periods alone are not enough to compare models with observations, as it is necessary to assess which simulated stars are undergoing DFMP. In linear models, this is easily determined using growth rates. In the nonlinear case, normally, the onset of DFMP corresponds to the radius $R_{\rm dom,0}$ at which dominant pulsation shifts from 1OM to FM. This parameter has already been determined for our model sequences, and in principle one should derive some analytic approximation of its dependence upon stellar and model parameters (similar to Eq.~10 of \citet{Trabucchi_etal_2017}). Instead, we only describe the mass dependence as
\begin{equation}\label{eq:onset_dom0}
    \log(R_{\rm dom,0}) = 2.130 + 1.150 \log(M/\Msun) - 0.496 \log(M/\Msun)^2 \,,
\end{equation}
which is depicted as a red line in Fig.~\ref{eq:onset_dom0}. There are three main motivations for this choice. Firstly, our aim is that of assessing the accuracy of nonlinear periods predictions, rather than investigating the onset of DFMP. Secondly, since $R_{\rm dom,0}$ depends upon turbulent viscosity, an intermediate step would be necessary involving the calibration of $\anu$, which is not trivial. Finally, linear models suggest that the onset of DFMP is sensitive to metallicity, a parameter not explored in the present study. These arguments justify the crude approximation given by Eq.~\ref{eq:onset_dom0}. Simulated stars whose current radius is larger than $R_{\rm dom,0}$ are assumed to undergo DFMP.

\begin{figure*}
    \includegraphics[width=.9\textwidth]{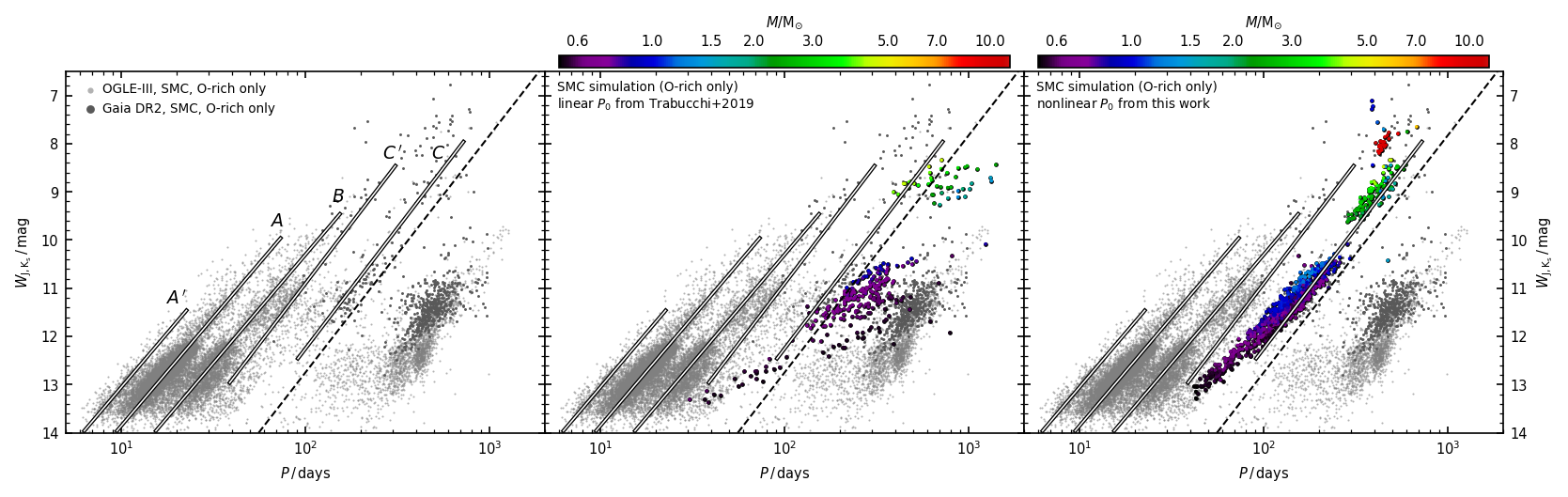}
    \includegraphics[width=.9\textwidth]{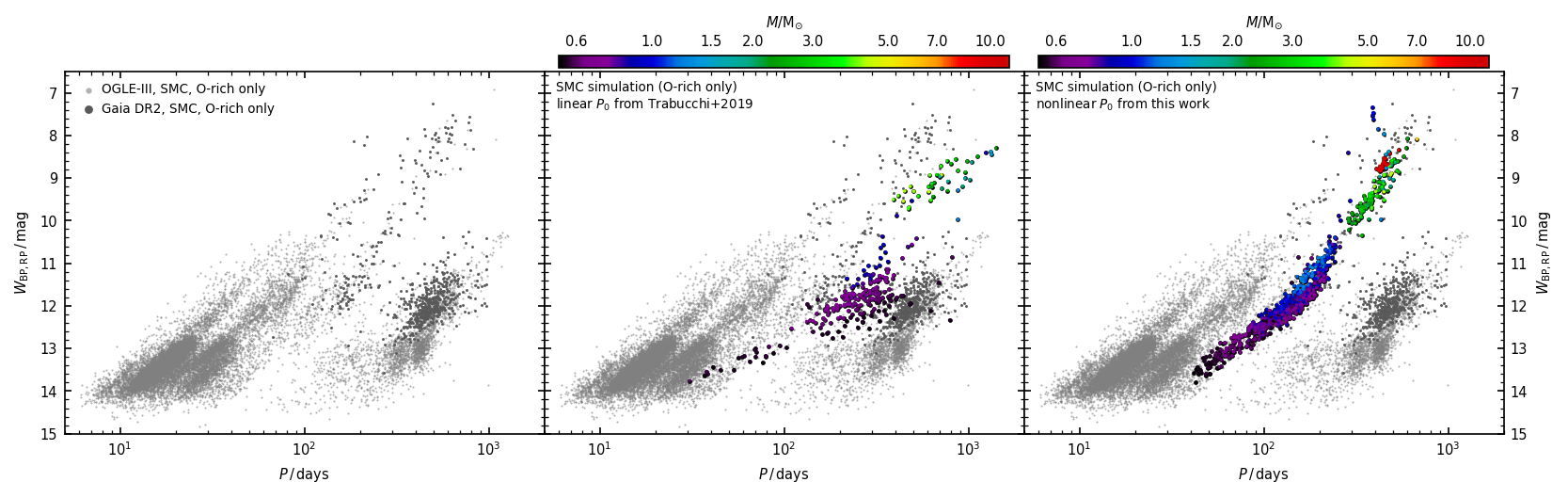}
    \caption{
    Period-luminosity diagrams of LPVs in the SMC using the Wesenheit indices $\wjk$ (top row) and $\wrp$ (bottom row). Small dots are observations are from OGLE-III (light grey dots) and \gaia\ DR2 (dark grey dots). The approximate location of PL sequences A$^{\prime}$, A, B, C$^{\prime}$, and C is indicated by solid lines, while dashed lines correspond to the ``death line'' displayed in Fig.~\ref{fig:DeathLine}. Simulated fundamental mode LPVs based on linear predictions from \citetalias{Trabucchi_etal_2019} or on nonlinear predictions from this work are shown, respectively, in panels on the central and right column. In both cases, they are colour-coded by current mass.
    }
    \label{fig:comparisonPLD_SMC}
\end{figure*}

\subsection{Period-luminosity diagrams}
\begin{figure*}
    \includegraphics[width=.9\textwidth]{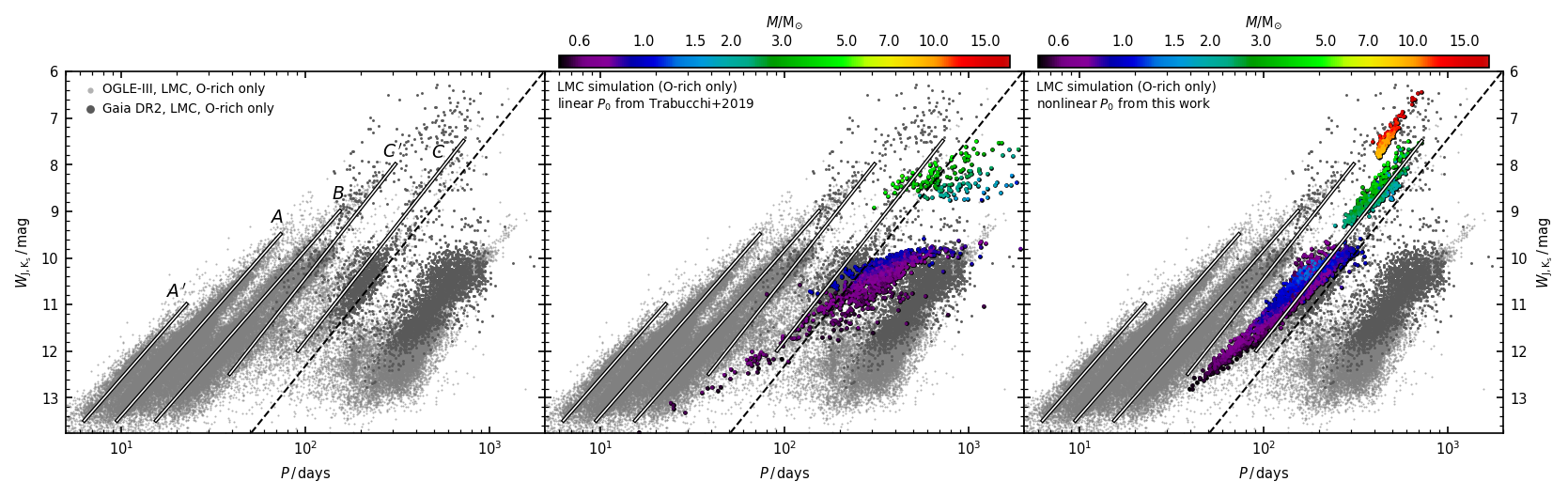}
    \includegraphics[width=.9\textwidth]{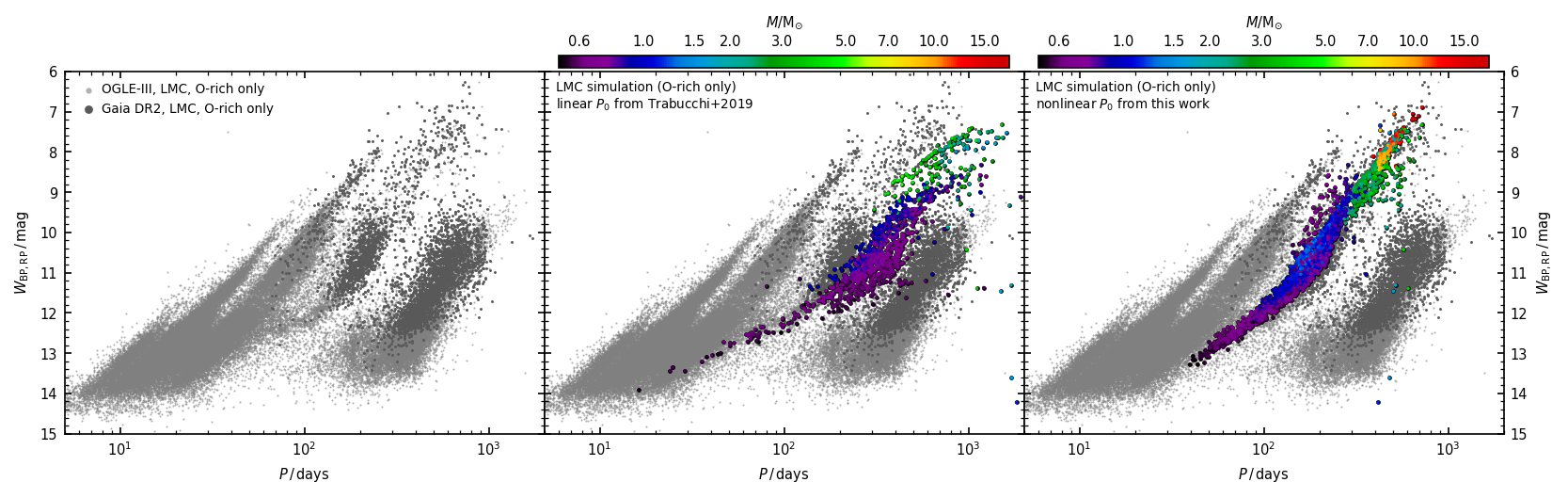}
    \caption{
    Similar to Fig.~\ref{fig:comparisonPLD_SMC}, but for the LMC.
    }
    \label{fig:comparisonPLD_LMC}
\end{figure*}

To construct period-luminosity diagrams (PLD), we make use of the Wesenheit indices \citep[see e.g.][]{Madore_1982,Lebzelter_etal_2019}
\begin{equation}\label{eq:wjk}
    \wjk = K_{\rm s} - 0.686 (J-K_{\rm s}) \,,
\end{equation}
\begin{equation}\label{eq:wrp}
    \wrp = \grp - 1.3 (\gbp - \grp) \,,
\end{equation}
which are obtained with 2MASS photometry in the $J$, $\ks$ filters and \gaia\ photometry in the $\gbp$, $\grp$ filters. The PLDs are shown in Fig.~\ref{fig:comparisonPLD_SMC} and~\ref{fig:comparisonPLD_LMC} for the SMC and LMC, respectively. Synthetic photometry for the simulation is obtained by means of tabulated bolometric corrections as explained in \citet{Pastorelli_etal_2019}.

Periods obtained from the newly derived fundamental mode PMR are in nice agreement with observations and show a substantial improvement with respect to linear predictions. Owing to the bending of the nonlinear PR relation, FM periods cannot increase arbitrarily past the observed PL sequence C, in contrast with what happens when linear models are employed. Despite being a crude approximation, the criterion determined from nonlinear models for the onset of DFMP also improves upon the approach based on linear growth rates. By capturing the earlier onset at low masses, it results in a more populated faint tail of the PL sequence, in better agreement with observations. The onset of DFMP is responsible for the left edge of PL sequence C at $\wjk\gtrsim8$, which we find to be fairly well reproduced by models.

The long-period edge of sequence C, on the other hand, is determined by the bending and saturation of the nonlinear PR relation. Eq.~\ref{eq:DeathLine} describing the ``death line'' in the PR diagram can be converted into a relation describing the maximum period of the FM at given $\wjk$. This is depicted by dashed lines in Fig.~\ref{fig:comparisonPLD_SMC} and~\ref{fig:comparisonPLD_LMC}, and reproduces remarkably well the slope and the right edge of PL sequence C. In this respect, PL sequence C is intrinsically different than other PL sequences, whose long-period edge is determined either by the shift of dominant pulsation to a lower-order mode, or by overtone pulsation becoming stable as the oscillation frequency becomes equal to the acoustic cut-off frequency at the surface \citepalias[][]{Trabucchi_etal_2019}. Note that stars shown in Figs.~\ref{fig:comparisonPLD_SMC} and~\ref{fig:comparisonPLD_LMC} are optically visible. Stars in the final high mass-loss rate superwind phase are detected as infrared sources only, and their periods can fall to the long-period side of the optical sequence C \citep[e.g.][]{Wood_2015}. These stars probably coincide with those we tried to model here but for which no stable limit cycle could be found due to the extremely large amplitude of pulsation (the models that were flagged in Fig.~\ref{fig:nonlinearPMRfit_M_alpha}).

Based on theoretical arguments, \citet{Wood_2015} showed how stellar mass decreases towards longer periods across PL sequences, at fixed luminosity. This effect, reproduced by our PMR relation, was first observed by \citet{Feast_etal_1989} and \citet{Hughes_Wood_1990} for Miras on the PL sequence C, and recently investigated in more detail by \citet{Lebzelter_etal_2019}. At $\wjk\simeq8.5$, mass decreases from $\sim4\,\Msun$ to $\sim2\,\Msun$ across PL sequence C. At fainter magnitudes this is less evident, as stars with $M\simeq0.8\,\Msun$ tend to be brought back towards the left edge of the PL sequence (see Sect.~\ref{ssec:evotracks}).
The offset is especially large for RSGs, predicted at $\wjk\simeq7.5$ in the region between PL sequences C$^{\prime}$ and C, in agreement with observations \citep{Lebzelter_etal_2019}. The location of RSGs is also determined by an earlier onset of DFMP at towards large masses (see Fig.~\ref{fig:nlP0_onset}).

\subsection{Evolutionary tracks}
\label{ssec:evotracks}
\begin{figure}
    \includegraphics[width=\columnwidth]{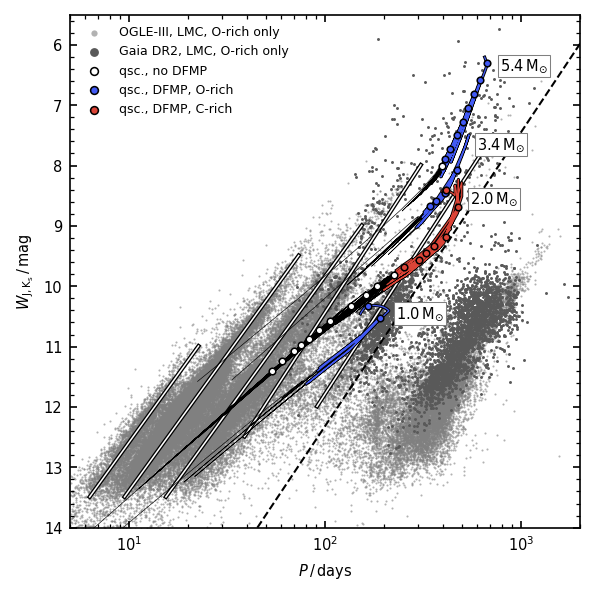}
    \caption{
    Period-luminosity diagram of LPVs in the the LMC in the form ($\wjk$, $P$). Curves are obtained from \colibri\ AGB evolutionary tracks, and represent the evolutionary path of the fundamental mode in the PLD for different initial masses (as labelled). Circles indicate stages of quiescent H-shell burning. Black lines and white circles correspond to stages at which the fundamental mode is not dominant. After it becomes dominant, lines and circles are coloured according to surface composition (blue: O-rich; red: C-rich). All tracks have initial metallicity $Z_{\rm i}=0.006$.
    }
    \label{fig:evoPLD_WJK_LMC}
\end{figure}

To better understand the implications of nonlinear PMR relation, it is instructive to follow the evolutionary path of FM pulsation in the period-luminosity diagram. To do so, we use Eq.~\ref{eq:nlPMR} to compute FM periods along AGB evolutionary tracks computed with the \colibri\ code. Bolometric luminosity is converted to magnitudes in the 2MASS passbands using \trilegal, and the result is displayed on top of observations in Fig.~\ref{fig:evoPLD_WJK_LMC}. We show tracks corresponding to initial metallicity $Z_{\rm i}=0.006$\footnote{
    Envelope metallicity increases during the TP-AGB, but it remains safely below $Z\simeq0.006$ during the O-rich phase for all evolutionary tracks used here, being thus consistent with the value used for pulsation models.
} and to a few different values of initial mass, $M_{\rm i}=1.0, 2.0, 3.4$, and $5.4\,\Msun$. At the beginning of each track, pulsation is linear, and the path followed by the FM period is less steep than the observed PL sequences, hence it crosses them. The PR relation bends after reaching the breaking radius $\Rb$, which is mirrored by a change of slope rather evident in the $5.4\,\Msun$ evolutionary track. At this point, the path in the PLD becomes steeper than the PL sequence. Note that a star can go back to the linear regime if it contracts after a thermal pulse.

If the tracks reach the saturation radius, the path becomes vertical, as seen in the brightest portion of the $2.0\,\Msun$ track. Note that the latter is representative of a star that becomes C-rich, hence the predicted periods are not necessarily realistic, even though they appear to be rather compatible with observations. This track shows rather well the fact that, due to the bending and saturation of the PR relation, a star that has crossed the PL sequence C and reached its long-period edge can be brought back near the left edge at brighter magnitudes.

\section{Summary and conclusions}
\label{sec:Conclusions}

We presented the results from the computation of a set of nonlinear pulsation models of O-rich AGB stars, widely covering the relevant range of stellar masses and luminosities. We found that nonlinear fundamental mode periods at large amplitudes are systematically shorter than in linear calculations, and are in good agreement with observations. The overtone mode periods were found to be entirely compatible with linear predictions. As long as the amplitude of pulsation (in terms of radial displacement of the optical surface) is not larger than $\sim50$ per cent of the mean radius, the nonlinear fundamental mode period is also consistent with predictions from linear models. At larger amplitudes, the linear approximation breaks down and the slope of the period-radius relation decreases abruptly with respect to the linear case. This is explained in terms of a substantial structural readjustment of the stellar envelope induced by large-amplitude pulsation. With respect to corresponding static models, the mean density increases in the pulsating layers, causing the period to decrease. At the largest radii, the fundamental mode period becomes independent of radius.

We modelled our results by means of an analytic period-mass-radius relation, that we tested against observations of long-period variables in the Magellanic Clouds using appropriate synthetic stellar population models. By capturing the earlier onset of dominant fundamental mode pulsation and its shallower dependence on stellar radius, nonlinear predictions result in much better agreement with observations than linear models. In particular, they are able to reproduce the observed PL sequence C, where Mira variables are found. The bending and saturation of the nonlinear period-radius relation is the very reason for the origin of the long-period edge of PL sequence C, that linear models fail to describe.

By delivering, for the first time, a way to accurately predict the period of fundamental mode long-period variables, this work represents an important achievement in the study of luminous red giant stars. The present models will help to fill the gap between theory and observation that has affected several subjects, from the study of mass loss on the AGB to the exploitation of Miras as standard candles. Our results will be especially important in view of the large wealth of stellar variability data expected from ongoing and future surveys such as \gaia, LSST, PLATO, and JWST.

A number of questions remain open, some of which were only briefly addressed in this work. In particular, our study suggests that chemical composition might have a significant impact on pulsation property, and requires systematic investigation. Pulsational stability as a function of global stellar parameters also needs to be better constrained, a task that would necessarily involve the investigation of multiperiodicity in nonlinear models. We explored the role of turbulent viscosity in linear and nonlinear pulsation, and found it to affect periods to a negligible level in most cases. In contrast, as expected, the amplitude and onset of pulsation are rather sensitive to the value employed for the turbulent viscosity parameter, a calibration of which is highly desirable. This is especially important as it represents a key step for the modelling of the light curves and photometric amplitudes of long-period variables. The ability to accurately predict such features would require great effort due to the necessity of modelling the complicated hydrodynamics of the atmospheric layers, including radiative transfer, chemistry and dust formation. At the same time, it would be highly rewarding, as it would provide additional tools to characterize in detail the properties of long-period variables and their hosting stellar populations.

\section*{Data availability}
The nonlinear time series generated in this research will made available on \url{http://starkey.astro.unipd.it/} and on VizieR. The catalogues of observed LPVs underlying this article are available from \citet{Soszynski_etal_2009_LMC,Soszynski_etal_2011_SMC} at \url{http://www.astrouw.edu.pl/ogle/ogle3/OIII-CVS/}, and from \citet{Mowlavi_etal_2018} at \url{https://gea.esac.esa.int/archive/}.

\section*{Acknowledgements}
M.T. and N.M. acknowledge the support provided by the Swiss National Science Foundation through grant Nr. 188697.
We acknowledge the support from the ERC Consolidator Grant funding scheme ({\em project STARKEY}, G.A. n. 615604).
This publication makes use of data products from the Two Micron All Sky Survey, which is a joint project of the University of Massachusetts and the Infrared Processing and Analysis Center/California Institute of Technology, funded by the National Aeronautics and Space Administration and the National Science Foundation.
This publication makes use of data from the \mbox{OGLE-III} Catalog of Variable Stars.
This research made use of \textsc{NumPy} \citep{numpy2020}, \textsc{SciPy} \citep{SciPy}, \textsc{matplotlib}, a Python library for publication quality graphics \citep{matplotlib}, and \textsc{Astropy}, a community-developed core Python package for Astronomy \citep{astropy2018}.



\bibliographystyle{mnras}
\bibliography{references}




\bsp	
\label{lastpage}
\end{document}